\newcommand\ezp{E_0^{+}(r)}
\newcommand\ezm{E_0^{-}(r)}

\newcommand\rp{r_+}
\newcommand\rmen{r_{-}}
\newcommand\rc{r_c}
\newcommand\rzer{r_0}
\newcommand\tigam{\tilde{\gamma}}
\newcommand\idq{{\mathbb I}_4}
\newcommand\idd{{\mathbb I}_2}
\newcommand\zed{{\mathbb O}_2}

\newcommand\pa{\partial}

\newcommand\beque{\begin{equation*}}
\newcommand\beq{\begin{equation}}
\newcommand\eeq{\end{equation}}
\newcommand\eeque{\end{equation*}}
\newcommand\beqnl{\begin{eqnarray}}
\newcommand\beqna{\begin{eqnarray*}}
\newcommand\eeqna{\end{eqnarray*}}
\newcommand\eeqnl{\end{eqnarray}}

\newcommand\RR{{\mathbb R}}
\newcommand\CC{{\mathbb C}}
\newcommand\ZZ{{\mathbb Z}}

\documentclass[aps,prd]{revtex4}
\usepackage{amsfonts}
\usepackage{amsthm}
\usepackage{amsbsy}
\usepackage{amssymb}
\usepackage{epsfig}
\usepackage{psfrag}

\begin{document}

\date{\today}

\title{Massive Dirac particles on the background of charged de-Sitter black hole manifolds}

\author{F. Belgiorno\footnote{E-mail address: belgiorno@mi.infn.it}}
\affiliation{Dipartimento di Fisica, Universit\`a di Milano, 20133
Milano, Italy, and\\
I.N.F.N., sezione di Milano, Italy}
\author{S. L. Cacciatori\footnote{E-mail address: sergio.cacciatori@uninsubria.it}}
\affiliation{Dipartimento di Fisica e Matematica, Universit\`a dell'Insubria, 22100
Como, Italy, and\\
I.N.F.N., sezione di Milano, Italy}

\begin{abstract}

We consider the behavior of massive Dirac fields on the background
of a charged de-Sitter black hole. All black hole geometries
are taken into account, including the Reissner-Nordstr\"{o}m-de-Sitter one,
the Nariai case and the ultracold case. Our focus is at first on the existence of bound
quantum mechanical states for the Dirac Hamiltonian on the given backgrounds.
In this respect, we show that in all cases no bound state is allowed, which
amounts also to the non-existence of normalizable time-periodic solutions
of the Dirac equation. This quantum result is in contrast to classical 
physics, and it is shown to hold true even for extremal cases. 
Furthermore, we shift our attention on the very interesting problem of the
quantum discharge of the black holes. Following Damour-Deruelle-Ruffini
approach, we show that the existence of level-crossing between
positive and negative continuous energy states is a signal of the
quantum instability leading to the discharge of the black hole, and
in the cases of the Nariai geometry and of the ultracold geometries we
also calculate in WKB approximation the transmission coefficient related to the 
discharge process.

\end{abstract}

\maketitle

\section{Introduction}
\label{intro}

Black holes in de Sitter space are an interesting subject of investigation, 
both on the theoretical side and on the experimental one. On one hand, 
the contextual presence of a black hole event horizon and of a cosmological event horizon, 
to be associated with the corresponding quantum emission of thermal radiation \cite{gibhaw} 
is a feature which enriches the framework of black hole thermodynamics in itself 
and also because of the possibility to obtain a true non-equilibrium situation 
when two different temperatures coexist on the same manifold. On the other hand, 
black hole physics in spacetimes with positive cosmological constant appear to be 
of direct physical interest, because the present-day measurements of cosmological 
parameters confirm the presence of a small positive cosmological constant, which implies that 
dS backgrounds are the real black hole backgrounds to be taken into account for 
physical considerations.\\
In this paper, we consider some relevant aspects of the physics of 
massive quantum Dirac particles on dS black hole backgrounds. 
We first show that, as expected, the Dirac Hamiltonian
is well behaved in the sense that its self-adjointness can be ensured 
without imposing any boundary condition. We also determine, 
by means of spectral analysis, two relevant physical properties: there 
is no mass gap in the spectrum, even if the particles are massive, and 
there exist no quantum bound state for charged particles around a charged 
black hole, in contrast to classical physics. The latter property amounts to the
absence of normalizable and time-periodic solutions of the
Dirac equation  on the background of a non-extremal Reissner-Nordstr\"{o}m-dS black hole,
in full agreement with the recent literature on
this topic \cite{finster-rn,finster-axi,yamada,batic,belcaccia}. Furthermore, we show that
this holds true also in the extremal case, due to the prominent role of the
cosmological event horizon, as well as in the so-called Nariai case and in the
ultracold ones.\\
In the second part of the paper, we also take into account the problem of
pair-creation by a charged black hole. This is a long-standing topic in the
framework of quantum effects in the field of a black hole, as old as the
Hawking effect but still different in its origin \cite{gibbons,khriplovich}. 
The latter can be brought back to vacuum instability in presence of 
an external field (see e.g. \cite{euler-heisenberg,schwinger} and, in the recent 
literature \cite{gitman,kim-page,kim}). 
It is shown that the
presence of level crossing, i.e. of overlap between positive continuum energy
states and negative continuum energy ones, according to a criterion introduced
by Ruffini, Damour and Deruelle \cite{chruffini,damo,deruelle,ruffini}, recently
extended to include the Reissner-Nordstr\"{o}m-AdS case \cite{belgcaccia},
is a still valid tool for investigating
pair-creation of charged Dirac particles even in presence of a positive
cosmological constant. We point out that this method is equivalent to the 
ones commonly exploited in order to investigate instability properties 
of the vacuum \cite{gitman,kim}, even if the criterion of level-crossing 
seems to be specific of the above references \cite{chruffini,damo,deruelle,ruffini}.\\ 
A special attention is focused on special cases, like
the Nariai and the ultracold ones, for which an estimate in WKB approximation 
of the transmission coefficient related to the process of pair-creation is provided.\\
This work, together with the analogous one concerning the Dirac equation on the
background of a Reissner-Nordstr\"{o}m-AdS black hole, completes the analysis
of the process of pair-creation by a charged black hole in presence of a
cosmological constant, and in this sense it also extends the analysis
on the background of a Reissner-Nordstr\"{o}m black hole \cite{gibbons,khriplovich}. 
We recall that, in spite of the fact that dS and AdS differ for a 
change of sign in the cosmological constant, very different manifolds and very 
different physics occur on these backgrounds. We mention for example the 
occurrence of closed timelike curves in the AdS case, 
a problem which can be 
overcome by passing to the universal covering, but at the price to deal with the 
lack of global hyperbolicity \cite{hawpage}. On the quantum level, self-adjointness of the 
wave operators cannot be ensured in general (see e.g. \cite{belgcaccia} for RN-AdS black holes), 
and boundary conditions have to be introduced for some cases, 
because of a boundary-like 
behavior of the AdS asymptotic region. In particular, for  $\mu \sqrt{\frac{3}{|\Lambda|}}< \frac 12$, 
where $\mu$ is the Dirac particle mass and $\Lambda$ is the (negative) cosmological constant, 
several boundary conditions can be chosen (see e.g. \cite{bachelot} for explicit choices of 
boundary conditions for the Dirac Hamiltonian on pure AdS), and then physics is not 
uniquely defined.\\
In the dS cases 
we discuss herein, no such features arise. Moreover, black 
holes are characterized by a single event horizon in the AdS case and by two 
event horizons in the dS one. This fact is shown to be at the root of the fact 
that in the de Sitter case there is always level-crossing, which is in contrast 
not only to the AdS case but also with the standard RN case ($\Lambda=0$). 
This feature is then peculiar of these solutions;  
notwithstanding, the actual presence of pair-creation is to
be associated with further conditions, to be related with the actual largeness of the
forbidden region separating positive energy states from negative energy ones.\\
For completeness, we recall that charged Dirac fields in the more general 
Kerr-Newman-de Sitter background 
have been studied with the aim to determine their quasinormal modes in \cite{konoplya}.

\section{Dirac Hamiltonian in the case $\rp<\rc$: Reissner-Nordstr\"{o}m-dS black holes}
\label{essauto}

We first define the one-particle Hamiltonian for
Dirac massive particles on the Reissner-Nordstr\"{o}m-dS
black hole geometry (RN-dS black hole in the following).
We use natural units ${\mathrm \hbar}={\mathrm c}={\mathrm G}=1$
and unrationalized electric units.
The metric of the RN-dS black hole manifold
$(t\in \RR;\ r\in (r_+,r_c);\ \Omega\in S^2)$ is
\beqnl
ds^2&=& -f(r) dt^2+\frac{1}{f(r)} dr^2 +r^2 d\Omega^2\cr
f(r)&=& 1-\frac{2 M}{r}+\frac{Q^2}{r^2}-\frac{\Lambda}{3} r^2;
\label{regeo}
\eeqnl
$M$ is the mass and $Q$ is the electric charge of the black hole, and
$\Lambda>0$ is the cosmological constant; let us define $L = \sqrt{\frac{3}{\Lambda}}$;
the equation $f(r)=0$ is assumed to admit solutions $r_c>\rp \geq \rmen >r_0$; then one obtains
\beq
f(r) = \frac{1}{L^2 r^2} (\rc -r)(r-\rp)(r-\rmen)(r-\rzer). 
\eeq
$r_c$ is the radius of the cosmological horizon,
$r_+$ is the radius of the black hole event horizon,
$\rmen$ is the radius of the Cauchy horizon. Moreover, due to the
actual lack of a term proportional to $r^3$, one has
$\rzer = -(\rc+\rp+\rmen)$. The above reparameterization of the metric
amounts to implementing the following relations between $\rc,\rp,\rmen$
and $M,Q,L$:
\beqna
L^2 &=& \rc (\rp+\rmen+\rc) +\rp^2+\rmen^2+\rp \rmen\cr
2 L^2 M &=& \rc^2 \rp + \rc \rp^2 + 2 \rc \rp \rmen + \rc^2 \rmen + \rc \rmen^2 + \rp^2 \rmen + \rp \rmen^2\cr
L^2 Q^2 &=& \rc \rp \rmen (\rc +\rp + \rmen).
\eeqna
It is not difficult to show that four real zeroes of $f(r)=0$ exist (and three are positive) if and only
if the following conditions are implemented:
\beqnl
&& 0 < Q^2 < \frac{L^2}{12} \\
&& M_{extr} \leq M < M_{max},
\label{bounds}
\eeqnl
where
\beq
 M_{extr} = \frac {L}{3\sqrt 6}\sqrt{ 1-\sqrt { 1-12 \frac {Q^2}{L^2} }}
\left( 2+\sqrt {1-12 \frac {Q^2}{L^2}}\right)
\label{mextr}
\eeq
is the mass of the extremal black hole with $\rmen = \rp$, and
\beq
 M_{max} = \frac {L}{3\sqrt 6}\sqrt{ 1+\sqrt { 1-12 \frac {Q^2}{L^2}}}
\left( 2-\sqrt {1-12 \frac {Q^2}{L^2}}\right)
\label{mmax}
\eeq
is the mass of the black hole with $\rp=\rc$ (see sect. \ref{sol-nariai}). See
\cite{belgcaccia-knds} for the analysis of the more general Kerr-Newman-dS case.

The vector potential associated with the
RN-dS solution is $A_{\mu}=(-Q/r,0,0,0)$. Spherical symmetry, as usual,
allows to separate variables \cite{soffel,wheeler,soffel1,thaller} and to obtain
the following reduced Hamiltonian
\beq
H_{red}=\left[
\begin{array}{cc}
-\sqrt{f}\; \mu+ \frac{e\; Q}{r} &\
f\; \partial_r +k\; \frac{\sqrt{f}}{r}\cr
-f\; \partial_r +k\; \frac{\sqrt{f}}{r}
&\  \sqrt{f}\; \mu+\frac{e\; Q}{r}
\end{array}
\right]
\label{reduced-ham}
\eeq
where  $f(r)$ is the same as in (\ref{regeo}),
$k=\pm (j+1/2) \in \ZZ-\{0\}$ is the
angular momentum eigenvalue and $\mu$ is the mass of the Dirac particle.  The Hilbert space in which
$H_{red}$ is formally defined
is the Hilbert space $L^2 [(\rp,\rc), 1/f(r)\; dr]^{2}$
of the two-dimensional
vector functions $\vec{g}\equiv \left(\begin{array}{c}
g_1 \cr
g_2 \end{array}\right)$ such that
$$
\int_{\rp}^{\rc}\; \frac{dr}{f(r)}\; (|g_1 (r)|^2+|g_2 (r)|^2)<\infty.
$$
As a domain for the minimal operator associated with
$H_{red}$ we can choose the following subset
of $L^2 [(r_+,\rc), 1/f(r)\; dr]^{2}$: the set
$C_{0}^{\infty}(r_+,\rc)^2$
of the two-dimensional vector functions
$\vec{g}$ whose components are smooth and of compact support
\cite{weidmann}.
It is useful to define a new tortoise-like variable $y$
\beq
\frac{dy}{dr}=\frac{1}{f(r)}
\eeq
and then one obtains $y\in \RR$, with
$y\to \infty \Leftrightarrow r\to \rc^-$ and
$y\to -\infty \Leftrightarrow r\to \rp^+$.
The reduced Hamiltonian becomes
\beq
H_{red}=D_{0}+V(y)
\label{red}
\eeq
where
\[ D_{0}=\left[
\begin{array}{cc}
0 &\   \partial_y \cr
-\partial_y &\ 0\end{array} \right]
\]
and
\[ V(r(y))=\left[
\begin{array}{cc}
-\sqrt{f}\; \mu+ \frac{e\; Q}{r} &\  k\; \frac{\sqrt{f}}{r} \cr
k\; \frac{\sqrt{f}}{r}
&\ \sqrt{f}\; \mu+\frac{e\; Q}{r}\end{array} \right].
\]
The Hilbert space of interest for the
Hamiltonian (\ref{red}) is $L^2 [\RR, dy]^{2}$.
We check if the one-particle Hamiltonian
is well-defined in the sense that no boundary conditions
are required in order to obtain a self-adjoint operator.
This means that we have to check if the reduced Hamiltonian is essentially self-adjoint;
with this aim, we check if the solutions of the equation
\beq
H_{red}\; \vec{g} =\lambda\; \vec{g}
\label{eigen}
\eeq
are square integrable in a right neighborhood of $y=-\infty$ and in
a left neighborhood of $y=+\infty$.
The so called Weyl alternative generalized to a system of first order
ordinary differential equations (\cite{weidmann}, theorem 5.6)
can be applied, in particular if  in a right neighborhood of $y=-\infty$
at least one solution not square
integrable exists for every $\lambda \in \CC$, then no
boundary condition is required and
the so-called limit point case (LPC) is verified; if instead
for every $\lambda\in \CC$ all the solutions of
$(H_{red}-\lambda) \vec{g}=0$ lie in $L^2[(-\infty,c), dy]^{2}$, with
$-\infty<c<\infty$, the so-called limit circle case (LCC) occurs
(and boundary conditions are required). Analogously one
studies the behavior of solutions in a left neighborhood of $y=\infty$.
The Hamiltonian operator is essentially self-adjoint if the LPC
is verified both at $y=-\infty$ and at $y=\infty$
(cf. \cite{weidmann}, theorem 5.7). In the case at hand, we can
refer to corollary to theorem 6.8 (p. 99) of \cite{weidmann}, both
for $y\to -\infty$ and for $y\to \infty$. Thus, the Dirac operator  defined on
$C_{0}^{\infty}(r_+,\rc)^2$
is essentially self-adjoint on the RN-dS black hole background.

\section{Qualitative spectral properties and time-periodic solutions in the case $\rp<\rc$.}

We first show that the essential spectrum of the unique self-adjoint extension of the Dirac Hamiltonian 
[still indicated with $H_{red}$] coincides with
$\RR$ both in the case of non-extremal black holes and in the extremal case. 
This feature is expected in presence of a black hole horizon 
and is well-known in the case of scalar particles \cite{kay-ergo}, and 
also verified in the case of Dirac particles on Kerr-Newman black hole 
manifold (see e.g. \cite{belg-rn,belgio}). We confirm that it is verified 
also in the present cases. From a physical point of view, it also implies that  
there is no room for isolated eigenvalues, and then that 
there is no ``standard'' bound state (in the sense that a charged particle with
charge opposite to the charge of the black hole cannot form a bound state which is
analogous to the bound state an electron forms around an atomic nucleus).
Moreover, a finer analysis allows also to conclude that, both in the non-extremal
case and in the extremal one, the point spectrum is empty, and then no quantum 
bound state, i.e. no possibility to obtain a
normalizable time-periodic solution of the Dirac equation exists.

\subsection{Essential spectrum}
\label{sub-essential}

One expects that, in presence of an event horizon, i.e. of a so-called ergosurface,
the mass gap vanishes and that the continuous spectrum includes the whole real line.
We recall that qualitative spectral methods for the Dirac equation (see e.g.
\cite{thaller,weidmann})
have been applied to Dirac fields
on a black hole background in \cite{belgio,yamada}.
In order to verify this property, we adopt the decomposition method \cite{weidmann}.
We split the interval $(\rp,\rc)$ at a inner point $r_1$ and then consider the formal
differential expression (\ref{reduced-ham}) restricted to the sub-intervals
$(\rp,r_1]$ and $[r_1,\rc)$. Roughly speaking, we refer to the aforementioned expressions as
to the ``restriction of the Hamiltonian $H_{red}$
to the interval $(\rp,r_1]$ and to the interval $[r_1,\rc)$'' and write e.g. $H_{red}|_{[r_1,\rc)}$
for the latter.
We limit ourselves to consider the latter restriction, which is relative to the novel feature
of space-time, with respect to previously discussed cases, represented by the cosmological horizon.
In the tortoise-like coordinate $y$ one finds a potential $P$ such that
\[ P=\left[
\begin{array}{cc}
-\sqrt{f}\; \mu+ \frac{e\; Q}{r} &\
k\; \frac{\sqrt{f}}{r}\cr
k\; \frac{\sqrt{f}}{r}
&\  \sqrt{f}\; \mu+\frac{e\; Q}{r}
\end{array}
\right]
\]
and it holds
\[ \lim_{y\to \infty} P(y) = P_0 =\left[
\begin{array}{cc}
\Phi_c  & 0\cr
0 & \Phi_c
\end{array}
\right]
\]
which is in diagonal form and whose eigenvalues coincide.
We apply theorem 16.6 p. 249 of Ref. \cite{weidmann}, which implies that, if
$\nu_{-},\nu_+$, with $\nu_{-}\leq \nu_+$, are the eigenvalues of the matrix $P_0$, then
$\{\RR-(\nu_{-},\nu_+)\} \subset \sigma_{e} (H_{red}|_{[y(r_1),\infty)})$ if
\beq
\lim_{y\to \infty} \frac{1}{y} \int_{\epsilon_0}^y dt |P(t)-P_0|=0,
\eeq
where $|\cdot|$ stays for any norm in the set of $2\times 2$ matrices (we choose the
Euclidean norm). In our case
one has to find the limit as $y\to \infty$ for the following expression:
\beq
\frac{1}{y} \int^{r(y)}_{r_1} dr \frac{1}{h(r)} \frac{1}{\sqrt{\rc-r}} \sqrt{2 \mu^2 h(r) +
2 \left( \Phi_+^2 (\rc-r)+k^2 h(r)\right) \frac{1}{r^2}},
\eeq
where we put $h(r)=\frac{f(r)}{\rc-r}$.
Both in the non-extremal case and in the extremal one, the above integral is finite as
$r\to \rc$ i.e. as $y\to \infty$, and then the limit is zero (we recall that the difference
between non-extremal case and extremal one from this point occurs when studying the
limit as $r\to \rp$, i.e. as $y\to -\infty$. In the extremal case $\rp=\rmen$ the
corresponding integral diverges but
a trivial use of the l'Hospital's rule allows to find that the aforementioned
limit is still zero).
As a consequence, we can state that
\beq
\sigma_{e} (H_{red})=\RR.
\eeq
A completely analogous conclusion can be stated for the restriction to $(-\infty,r_1)$,
and again the essential spectrum contribution one finds is $\RR$ both in the non-extremal
case and in the extremal one.

\subsection{Absence of states of the point spectrum}
\label{ac-spec}

Qualitative spectral analysis of the reduced Hamiltonian in the non-extremal case can
be implemented by means of theorems in \cite{hs-ac} or also in \cite{weidmann}.
In \cite{belgcaccia-knds} a proof was given for the more general case of
the Dirac equation in a Kerr-Newman-de Sitter black hole background, again in the case
$\rp<\rc$. For the sake of completeness, we sketch the strategy and also provide
some details involving some differences with respect to \cite{belgcaccia-knds}.

We note that, given a decomposition point $r_1 \in (r_{+},r_c)$, we can introduce the following
self-adjoint operators $H_{hor}$ and $ H_{c}$ on the respective domains
$D(H_{hor})=\{ \vec{g}\in L^2 [(r_{+},r_1), 1/f(r)\; dr]^{2},\; \vec{g}
\hbox{ is locally absolutely continuous}; g_1 (r_1)=0;\;
H_{hor} \vec{g} \in L^2 [(r_{+},r_1), 1/f(r)\; dr]^{2}\}$, and analogously
$D(H_{c})=\{ \vec{g}\in L^2 [(r_{1},r_c), 1/f(r)\; dr]^{2},\; \vec{g}
\hbox{ is locally absolutely continuous}; g_1 (r_1)=0;\;
H_{c} \vec{g} \in L^2 [(r_{1},r_c), 1/f(r)\; dr]^{2}\}$. According to the 
decomposition method applied to the absolutely continuous spectrum, 
one has $\sigma_{ac} (H_{red}) = \sigma_{ac} (H_{hor})\cup \sigma_{ac} (H_{c})$ (cf. e.g. 
\cite{belgcaccia-knds}). Theorem 16.7 in \cite{weidmann}
allows to conclude that, in the non-extremal case, $H_{hor}$ has absolutely continuous 
spectrum in $\RR-{\Phi_+}$, where $\frac{\Phi_+}{e}$ is the electrostatic
potential at the black hole event horizon, and that $H_{c}$ has 
absolutely continuous spectrum in $\RR-{\Phi_c}$, where $\frac{\Phi_c}{e}$ is the electrostatic
potential at the cosmological horizon. Moreover, for $e Q>0$ it has to hold $\Phi_c<\Phi_+$,
and $\Phi_c>\Phi_+$ for $e Q<0$, due to the inequality $\rc>\rp$. In any case,
$\Phi_c\ne \Phi_+$ occurs, and this is an interesting fact in the light of the
study of the pair-creation process, as we shall see in the following section.
As to the spectral properties of the reduced Hamiltonian, one can easily infer that
the spectrum is absolutely continuous in $\RR$ (indeed, the above analysis
allows to conclude that the spectrum is absolutely continuous in
$\RR - \{\{\Phi_c\}\cap \{\Phi_+\}\}$ but of course the latter set coincides with $\RR$).\\
As to the extremal case, one can again refer to theorem 16.7 in \cite{weidmann} for $H_{c}$
and to theorem 1 in \cite{hs-ac}  for $H_{hor}$ to conclude that the spectrum is absolutely continuous in
$\RR - \{\{\Phi_c\}\cap \{\Phi_+\}\}$. Again, the latter set is $\RR$.

\section{Nariai solution}
\label{sol-nariai}

We take into consideration the special case of the so-called charged Nariai
solution  \cite{mann,bousso}, which is a black hole solution with $\rmen<\rp=\rc$. As known, the metric
(\ref{regeo}) is no more valid and a suitable transformation is necessary. It
can be shown that the manifold can be described by
\beq
ds^2 = \frac{1}{A} (-\sin^2 (\chi) d \psi^2 + d\chi^2) + \frac{1}{B} (d\theta^2+
\sin^2 (\theta) d\phi^2)
\label{nariai}
\eeq
with $\psi \in \RR, \chi\in (0,\pi)$ and where
$B=\frac{1}{2 Q^2}\left( 1-\sqrt{1-12 \frac{Q^2}{L^2}} \right)$ and $A=\frac{6}{L^2}-B$ are constants
such that $\frac{A}{B}<1$ \cite{mann,bousso}. We note that there is no warping factor in
the metric between the ``radial'' part and the $S^2$ part. For an electrically charged
black hole we can choose $A_{i} = -Q \frac{B}{A} \cos (\chi) \delta_{i}^0$.\\
We study the Dirac equation as in \cite{soffel,wheeler}. With the same notation
as in \cite{soffel}, we introduce the so-called generalized Dirac matrices such that
$\{ \gamma_i, \gamma_j \} = 2 g_{ij}$:
\beqnl
&&\gamma_0 = \frac{\sin (\chi)}{\sqrt{A}} \tigam_0 \quad \quad  \gamma^0 = -\frac{\sqrt{A}}{\sin (\chi)}  \tigam_0 \cr
&&\gamma_1 = \frac{1}{\sqrt{A}} \tigam_1 \quad \quad \hphantom{....} \gamma^1 = \sqrt{A}  \tigam_1 \cr
&&\gamma_2 = \frac{1}{\sqrt{B}} \tigam_2 \quad \quad \hphantom{....} \gamma^2 = \sqrt{B}  \tigam_2 \cr
&&\gamma_3 = \frac{\sin (\theta)}{\sqrt{B}} \tigam_3 \quad \quad \hphantom{.}\gamma^3 = \frac{\sqrt{B}}{\sin (\theta)}  \tigam_3,
\eeqnl
where $\tigam_i$, $i=0,1,2,3$ are the usual Dirac matrices in Minkowski space. The Dirac equation is
\beq
[\gamma^k (\partial_k - \Gamma_k) - \mu]\Psi =0,
\eeq
with
\beq
\Gamma_k = -\frac{1}{4} \gamma^j (\partial_k \gamma_j -\gamma_l \Gamma_{jk}^l ) + i e A_k.
\eeq
One finds the following non-vanishing Christoffel symbols $\Gamma_{01}^0 = \cot (\chi),
\Gamma_{00}^1 = \sin (\chi) \cos (\chi), \Gamma_{33}^2 = -\sin (\theta) \cos (\theta),
\Gamma_{23}^3 = \cot (\theta)$. Then, due to our choice for $A_{i}$, we get
$\Gamma_0 =-\frac 12 \cos (\chi) \tigam_0 \tigam_1 + i e A_0, \Gamma_1 = 0, \Gamma_2 = 0,
\Gamma_3 = \frac{1}{2} \cos (\theta) \tigam_2 \tigam_3$. Then the Dirac equation becomes
\beqnl
&&\left[ -\frac{\sqrt{A}}{\sin (\chi)}  \tigam_0 (\partial_{\psi} - i e A_0) +
\sqrt{A}  \tigam_1 (\partial_{\chi} +\frac 12 \cot (\chi))+ \right. \cr
&& \left.
\sqrt{B} \left( \tigam_2 \left( \partial_{\theta} + \frac 12 \cot (\theta) \right)+ \tigam_3
\frac{1}{\sin (\theta)}  \partial_{\phi} \right) - \mu \right] \Psi = 0.
\eeqnl
By posing $\Psi = (\sin (\chi))^{-1/2} (\sin (\theta))^{-1/2} \zeta$, we eliminate the terms proportional to $\cot (\theta)$
and to $\cot(\chi)$
in the previous equation. We now consider a static solution with $\zeta = \exp (-i \omega \psi)
\eta (\chi,\theta,\phi)$. Then a trivial manipulation of the Dirac equation leads to the following
eigenvalue equation:
\beq
H \eta = \omega \eta,
\eeq
with
\beq
H = -i \tigam_0 \tigam_1 \sin (\chi) \partial_{\chi} - e A_0 \idq 
+ \sqrt{\frac{B}{A}} \sin (\chi) \tigam_1 K +i \tigam_0\frac{\mu}{\sqrt{A}} \sin (\chi).
\label{nariai-ham}
\eeq
$\idq$ stays for the $4\times 4$ identity matrix and $K$ is the following
operator:
\beq
K = -i \tigam_1 \tigam_0 \tigam_2 \partial_{\theta}  -i \tigam_1 \tigam_0
\tigam_3 \frac{1}{\sin (\theta)}  \partial_{\phi}
\eeq
which commutes with $H$ and whose eigenvalues are $k\in \ZZ-{0}$ \cite{soffel,wheeler}. By restricting $H$ to eigenspaces of
$K$ and by choosing
$$
\tigam_0 =
\left(
\begin{array}{cc}
i \idd & \zed \cr
\zed & -i \idd
\end{array}
\right), \quad \quad
\tigam_1 =
\left(
\begin{array}{cc}
\zed & \idd \cr
\idd & \zed
\end{array}
\right)
$$
($\idd$ is the $2\times 2$ identity matrix, $\zed$ is the $2\times 2$ zero matrix), we obtain
the reduced Hamiltonian
\beqnl
H_k &=& -\sin(\chi) \pa_{\chi} \left(
\begin{array}{cc}
\zed & -  \idd \cr
\idd & \zed
\end{array}
\right)  + e Q \frac{B}{A} \cos (\chi) \left(
\begin{array}{cc}
\idd & \zed \cr
\zed & \idd
\end{array}
\right) \cr
&+& \sqrt{\frac{B}{A}} \sin (\chi) k \left(
\begin{array}{cc}
\zed & \idd \cr
\idd & \zed
\end{array}
\right) - \frac{\mu}{\sqrt{A}} \sin (\chi) \left(
\begin{array}{cc}
\idd & \zed \cr
\zed & -\idd
\end{array}
\right) \cr
& = & h_k \otimes \idd,
\eeqnl
where
\beq
h_k = \left[
\begin{array}{cc}
e Q  \frac{B}{A} \cos (\chi) - \frac{\mu}{\sqrt{A}} \sin (\chi)
& \sin(\chi) \pa_{\chi} + \sqrt{\frac{B}{A}} \sin (\chi) k \cr
-\sin(\chi) \pa_{\chi} + \sqrt{\frac{B}{A}} \sin (\chi) k
& e Q  \frac{B}{A} \cos (\chi) + \frac{\mu}{\sqrt{A}} \sin (\chi)
\end{array}
\right]
\label{hk-nariai}
\eeq
The coordinate transformation
\beq
x = \log (\tan (\frac{\chi}{2})) \longleftrightarrow \chi = 2 \arctan (\exp (x))
\label{x-nariai}
\eeq
is such that $x\in \RR$ and, furthermore, $h_k$ becomes
\beq
h_k = \left[
\begin{array}{cc}
0 & \pa_x \cr
- \pa_x & 0
\end{array}
\right]+P(\chi(x)),
\label{hamiltonian-nariai}
\eeq
where
\beq
P(\chi) = \left[
\begin{array}{cc}
e Q  \frac{B}{A} \cos (\chi) - \frac{\mu}{\sqrt{A}} \sin (\chi)
& \sqrt{\frac{B}{A}} \sin (\chi) k \cr
  \sqrt{\frac{B}{A}} \sin (\chi) k
& e Q  \frac{B}{A} \cos (\chi) + \frac{\mu}{\sqrt{A}} \sin (\chi)
\end{array}
\right].
\label{nariai-pot}
\eeq
$h_k$ is formally self-adjoint in $L^2 [\RR, dx]^2$ and it is essentially self-adjoint
in $C_0^{\infty} (\RR)^2$, as follows from corollary to theorem 6.8 (p. 99) of \cite{weidmann}
(the limit point case occurs both at $x=-\infty$ and at $x=\infty$). It is easy to
show that the essential spectrum of $h_k$ coincides with $\RR$ and the same is true for the
absolutely continuous spectrum. The latter claim can be checked by following the ideas
displayed in sect. \ref{ac-spec}. See Appendix \ref{spe-nariai} for more details.

\section{Ultracold case}
\label{ultracold}

There is still a sub-case to be taken into account. It corresponds to the
so-called ultracold case \cite{mann}, where the three horizons coincide: $\rmen=\rp=\rc$.
Also in this case the metric (\ref{regeo}) is no more valid, and a suitable
limit has to be considered \cite{mann}. Actually, one can introduce two different metrics
for the ultracold case. As a consequence, also our analysis is split into two parts.

\subsection{ultracold I}

A first metric \cite{mann} is
\beq
ds^2 = -\chi^2 d \psi^2 + d\chi^2+ \frac{1}{2\Lambda} (d\theta^2+
\sin^2 (\theta) d\phi^2),
\label{ultracold-I}
\eeq
with $\chi\in (0,\infty)$ and $\psi\in \RR$.
One gets $\Gamma_{01}^0 = \frac{1}{\chi},
\Gamma_{00}^1 = \chi, \Gamma_{33}^2 = -\sin (\theta) \cos (\theta),
\Gamma_{23}^3 = \cot (\theta)$.
The electromagnetic field strength is $F=\sqrt{\Lambda} \chi d\chi\wedge d\psi$, and we
can choose $A_0 = \frac{\sqrt{\Lambda}}{2} \chi^2$ and $A_j=0$, $j=1,2,3$.
We introduce
\beqnl
&&\gamma_0 = \chi \tigam_0 \quad \quad \quad \quad \gamma^0 = -\frac{1}{\chi}  \tigam_0 \cr
&&\gamma_1 = \tigam_1 \quad \quad \quad \hphantom{...G} \gamma^1 = \tigam_1 \cr
&&\gamma_2 = \frac{1}{\sqrt{2\Lambda}} \tigam_2 \quad \quad \hphantom{o}
\gamma^2 = \sqrt{2 \Lambda}  \tigam_2 \cr
&&\gamma_3 = \frac{\sin (\theta)}{\sqrt{2\Lambda}} \tigam_3
\quad \quad \hphantom{}\gamma^3 = \frac{\sqrt{2\Lambda}}{\sin (\theta)}  \tigam_3,
\eeqnl
and then we obtain
$\Gamma_0 =-\frac 12 \tigam_0 \tigam_1 + i e A_0, \Gamma_1 = 0, \Gamma_2 = 0,
\Gamma_3 = \frac{1}{2} \cos (\theta) \tigam_2 \tigam_3$.
Calculations which are strictly analogous to the ones performed in the Nariai case
(with $\Psi = \exp (-i \omega \psi) \frac{1}{\sqrt{\chi} \sqrt{\sin (\theta)}} \zeta$)
and the variable change $\chi = \exp (x)$, $x\in \RR$,
lead to the following reduced Hamiltonian:
\beq
h_k = \left[
\begin{array}{cc}
-\frac{e \sqrt{\Lambda}}{2} \exp (2 x) -\mu \exp (x)
& \pa_{x} +\sqrt{2 \Lambda} k \exp (x)\cr
-\pa_{x} + \sqrt{2 \Lambda} k \exp (x)
&-\frac{e \sqrt{\Lambda}}{2} \exp (2 x) +\mu \exp (x)
\end{array}
\right].
\label{hk-ucI}
\eeq
As in the Nariai case,
$h_k$ is formally self-adjoint in $L^2 [\RR, dx]^2$ and it is essentially self-adjoint
in $C_0^{\infty} (\RR)^2$, as follows from corollary to theorem 6.8 (p. 99) of \cite{weidmann}.
The analysis of the spectrum can be pursued by means of the decomposition method applied to
the absolutely continuous spectrum, and one can again conclude that the absolutely continuous
spectrum of the self-adjoint extension of $h_k$ on $\RR$
coincides with the whole real line.  See Appendix \ref{spe-uI} for details.

\subsection{ultracold II}

The second allowed metric \cite{mann} for the ultracold case is
\beq
ds^2 = -d \psi^2 + dx^2+ \frac{1}{2\Lambda} (d\theta^2+
\sin^2 (\theta) d\phi^2),
\label{ultracold-II}
\eeq
with $x\in \RR$ and $\psi\in \RR$.
One gets $\Gamma_{33}^2 = -\sin (\theta) \cos (\theta),
\Gamma_{23}^3 = \cot (\theta)$.
The electromagnetic field strength is $F=-\sqrt{\Lambda}  d\psi\wedge dx$, and we
can choose $A_0 = \sqrt{\Lambda} x$ and $A_j=0$, $j=1,2,3$.
We introduce
\beqnl
&&\gamma_0 = \tigam_0 \quad \quad \quad  \hphantom{......G} \gamma^0 = -\tigam_0 \cr
&&\gamma_1 = \tigam_1 \quad \quad \quad \hphantom{......G} \gamma^1 = \tigam_1 \cr
&&\gamma_2 = \frac{1}{\sqrt{2\Lambda}} \tigam_2 \quad \quad \hphantom{....}
\gamma^2 = \sqrt{2 \Lambda}  \tigam_2 \cr
&&\gamma_3 = \frac{\sin (\theta)}{\sqrt{2\Lambda}} \tigam_3
\quad \quad \hphantom{..}\gamma^3 = \frac{\sqrt{2\Lambda}}{\sin (\theta)}  \tigam_3,
\eeqnl
and then we obtain
$\Gamma_0 =i e A_0, \Gamma_1 = 0, \Gamma_2 = 0,
\Gamma_3 = \frac{1}{2} \cos (\theta) \tigam_2 \tigam_3$.
Again calculations as above
(with $\Psi = \exp (-i \omega \psi) \frac{1}{\sqrt{\sin (\theta)}} \zeta$)
lead to the following reduced Hamiltonian:
\beq
h_k = \left[
\begin{array}{cc}
-e \sqrt{\Lambda} x -\mu
& \pa_{x} +\sqrt{2 \Lambda} k \cr
-\pa_{x} + \sqrt{2 \Lambda} k
&-e \sqrt{\Lambda} x +\mu
\end{array}
\right].
\label{hk-ucII}
\eeq
Also in this case,
$h_k$ is formally self-adjoint in $L^2 [\RR, dx]^2$ and it is essentially self-adjoint
in $C_0^{\infty} (\RR)^2$. As in the previous case, the decomposition method applied to
the absolutely continuous spectrum allows to draw the conclusion that the spectrum
of the self-adjoint extension of the Hamiltonian (\ref{hk-ucII}) is absolutely continuous
and coincides with $\RR$.
See Appendix \ref{spe-uII} for details.

\section{Pair Creation and Level-Crossing}

We follow the Ruffini-Damour-Deruelle \cite{chruffini,damo,deruelle,ruffini} approach, which 
was summarized in a previous paper \cite{belgcaccia}. Herein, we limit ourselves to recall
some very basic properties, focusing on the RN-dS case (the other cases can be dealt with 
analogously). In this approach one introduces effective potentials
$E_0^{\pm}(r)$ for the positive and negative
energy states respectively; they represent the classical turning points
for the particle motion and lead to the definition of the so-called effective
ergosphere.
These potentials enter the Hamilton-Jacobi (HJ) equation for a
classical particle. They can be interpreted also at the quantum level, as
in \cite{ruffini}. In particular, they indicate the regions
of level-crossing between positive and negative energy states
\cite{damo,deruelle}. In the case of the Dirac equation, it is known
that the HJ equation corresponds to a WKB approximation to the
Dirac equation at the lowest order \cite{pauli,rubinow}.
Variable separation in the quantum case allows to obtain an obvious improvement
of the semi-classical formulas, amounting in replacing the classical value of the
angular momentum with the quantum eigenvalues of the corresponding quantum operator \cite{belgcaccia}.
We limit ourselves to recall one of the main point of our analysis of the
Dirac Hamiltonian in \cite{belgcaccia}. The key-observation resides in the following
fact: if one consider the potential term in the Dirac Hamiltonian
\[ V(r)=\left[
\begin{array}{cc}
p_{11}(r) &  p_{12}(r) \cr
p_{21}(r)
& p_{22}(r)\end{array} \right],
\]
and formally calculates the eigenvalues of the above matrix, which are
found by solving
\beq
(p_{11}(r)-\lambda)(p_{22}(r)-\lambda)-p_{12}(r) p_{21}(r)=0;
\eeq
then, defining $S(r)\equiv \sqrt{(p_{11}(r)+p_{22}(r))^2-4 p_{11}(r) p_{22}(r)
+4 p_{12}(r) p_{21}(r)}$ one finds
\beq
\lambda^{\pm} (r)=\frac{1}{2}\left(
p_{11}(r)+p_{22}(r)\pm S(r)\right),
\eeq
and moreover one gets
\beq
\lambda^{\pm} (r)=E_0^{\pm}(r),
\eeq
i.e., the semi-classical potentials introduced by Damour-Deruelle-Ruffini
coincide with the eigenvalues of the matrix
potential term in the Dirac Hamiltonian.
Moreover, the square of the classical angular momentum term is replaced by the
square of the eigenvalues $k=\pm (j+1/2)$ for the quantum angular momentum, and
one obtains
\beq
E_0^{\pm}(r)=\frac{eQ}{r}\pm \sqrt{f(r) (\mu^2+\frac{k^2}{r^2})}.
\eeq

\subsection{Level crossing in the RN-dS case}

Level-crossing amounts to the presence of overlap between the range of
$E_0^{+}$ and the range of $E_0^{-}$, signalling the possibility of a tunneling
between positive energy states and negative energy ones; the latter phenomenon
is in turn interpreted as pair-creation at the barrier potential, and is
strictly related to the Klein paradox (incidentally, it could be called for this
reason also Klein effect \cite{belgio}). There is a peculiar property in the case of RN-dS
black holes: due to the inequality  $\Phi_c\ne \Phi_+$, an overlap is always
present, indeed one gets
\beq
\lim_{r\to \rp} E_0^{\pm}(r)=\Phi_+,
\eeq
and
\beq
\lim_{r\to \rc} E_0^{\pm}(r)=\Phi_c.
\eeq
Assuming for definiteness $e Q>0$, one finds
\beq
E_0^{-}(\rp)=\Phi_+ > \Phi_c = E_0^{+}(\rc),
\eeq
which proves the above claim. Moreover, it is easy to realize that level crossing 
occurs for energy $\omega$ such that 
\beq
\Phi_{c}=\min E_0^{+}(r) \leq \omega \leq \max E_0^{-}(r)=\Phi_+. 
\label{lev-cross}
\eeq
For $e Q<0$ the overlap still exists, being 
$E_0^{-}(\rc) = \Phi_c > \Phi_+ = E_0^{+}(\rp)$, and (\ref{lev-cross}) changes accordingly: 
$\Phi_{+} \leq \omega \leq \Phi_c$.\\
It is to be immediately pointed out that
this overlap is not enough to conclude that a sensible pair-creation
process occurs in the given background. Indeed, the potential barrier to be
overcome by the negative energy particle can be as large as the whole external
region of the spacetime. As a consequence, no efficient process can be expected on
this ground in general, and further conditions taking into account the effective
largeness of the barrier have to be looked for. Notice also that $E_0^+$, as a
function of $\mu^2$, is increasing, and the same is true for its dependence on $k^2$;
both these properties are reversed in the case of $E_0^-$, which is in fact decreasing.
As a consequence, one expects a more difficult level-crossing for increasing values
of $\mu^2$ or $k^2$.\\
We start giving sample-examples for a non-extremal black hole manifold. In figure 1,
we keep fixed the geometric background parameters $L,M,Q$ and the charge $e$ of the
Dirac particle, and consider two sample values of the mass. On the left,
we find the former phenomenon described above, i.e. a level-crossing
with a barrier as large as $\sim \rc-\rp$. On the right-hand case, which is obtained
by considering a lower fermion mass, we obtain instead a level-crossing associated
with a much smaller extent of the barrier. The latter case is expected to be
involved in a effective phenomenon of pair creation at the barrier. In figures 2 and 3
a further non-extremal case is displayed, with $\frac{\rc}{\rp}\sim 1.01$ (to be compared
with $\frac{\rc}{\rp}\sim 10.4$ of the example displayed in figure 1). In figure 3
we show the details of the potentials
near the horizons. In figure 4, an extremal case is also shown. In figure 5 we display the
so-called lukewarm case \cite{romans,mann}, 
which is such that the same temperature occurs in the case of the cosmological
horizon and of the black hole event horizon, still with $\rp<\rc$ (this happens
for $Q=M$). It is displayed for the sake of completeness, even if
from the point of view of the given phenomenon no peculiar behavior is expected with
respect to the cases explored in figures 1 and 2.\\
Notice that,
from a physical point of view, the phenomenon of pair creation by a charged black hole has been
related to the Schwinger calculation of pair creation by a homogeneous electric field
\cite{gibbons}. The highest value of the electrostatic potential, which
corresponds to the highest intensity of the electrostatic field, occurs near the
black hole event horizon, hence one could naively expect that the standard condition
$\Phi_+^2>\mu^2$, i.e.
\beq
\left(\frac{Q}{\rp}\right)^2 > \frac{\mu^2}{e^2},
\label{standard}
\eeq
which is enough for the standard Reissner-Nordstr\"{o}m case \cite{gibbons,khriplovich,damo}, is also
qualitatively relevant in the present one, at least in some approximation. We recall
that, in the case of the lightest known charged particle, i.e. the electron,
one has $\frac{\mu_e}{e} \sim 10^{-21}$. The above condition is not necessary for
the existence of a sensible level-crossing. A proof of its sufficiency in the
extremal case can be given under the hypothesis $\frac{k^2}{\rp^2}\ll 1$,
which allows to neglect the angular momentum contribution in the potentials,
and also for $\rp\ll \rc$. In the extremal case, it is easy to show that the condition to be
satisfied is that $\frac{dE_0^+}{dr}(\rp) \frac{dE_0^-}{dr}(\rp)>0$, which means
that both potentials are increasing or decreasing near $r=\rp$. One has to impose
\beq
\frac{dE_0^+}{dr}(\rp)\; \frac{dE_0^-}{dr}(\rp)=\frac{1}{\rp^2} \left[
\Phi_+^2 - \frac{\rc^2}{L^2}
\left(1-\frac{\rp}{\rc}\right) \left(1+3\frac{\rp}{\rc} \right)
\left(\mu^2+\frac{k^2}{\rp^2}\right) \right] >0,
\eeq
and then, under the above hypotheses $\frac{k^2}{\rp^2}\ll 1$ and $\rp\ll \rc$, one obtains
\beq
\frac{dE_0^+}{dr}(\rp)\; \frac{dE_0^-}{dr}(\rp)\simeq
\frac{1}{\rp^2} \left(\Phi_+^2 -\mu^2 \frac{\rc^2}{L^2} \right)>0,
\eeq
for which, being $\frac{\rc}{L}<1$, the aforementioned condition (\ref{standard})
is sufficient. In figure 6 we show an example of extremal black hole where a significant level crossing
occurs but condition (\ref{standard}) is not satisfied.

\begin{figure}[htbp!]
\begin{picture}(150,150)
\put(-140,160){\includegraphics[height=.30\textheight,
                      width=.24\textheight,angle=-90]{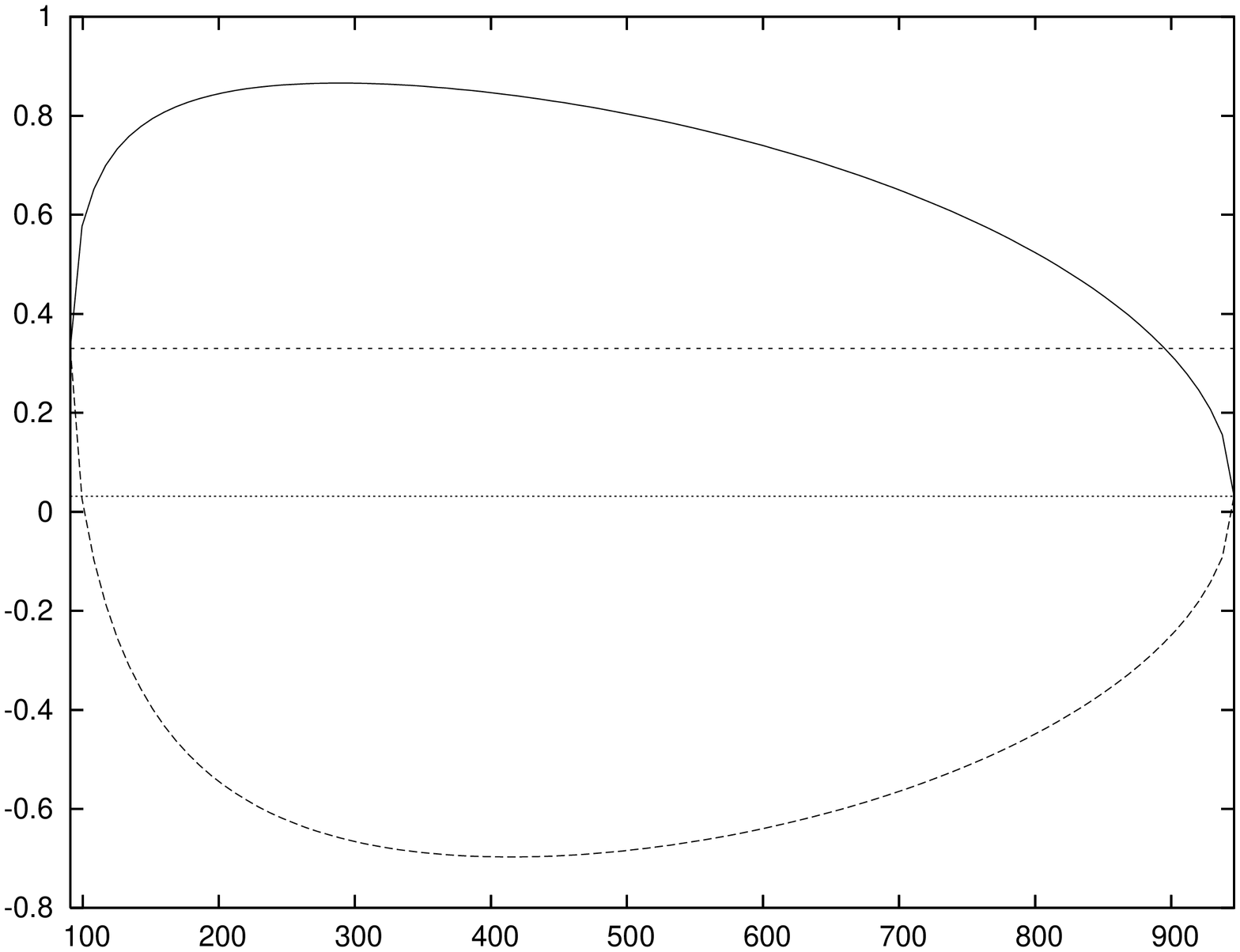}} 
\put(90,160){\includegraphics[height=.30\textheight,
                      width=.24\textheight,angle=-90]{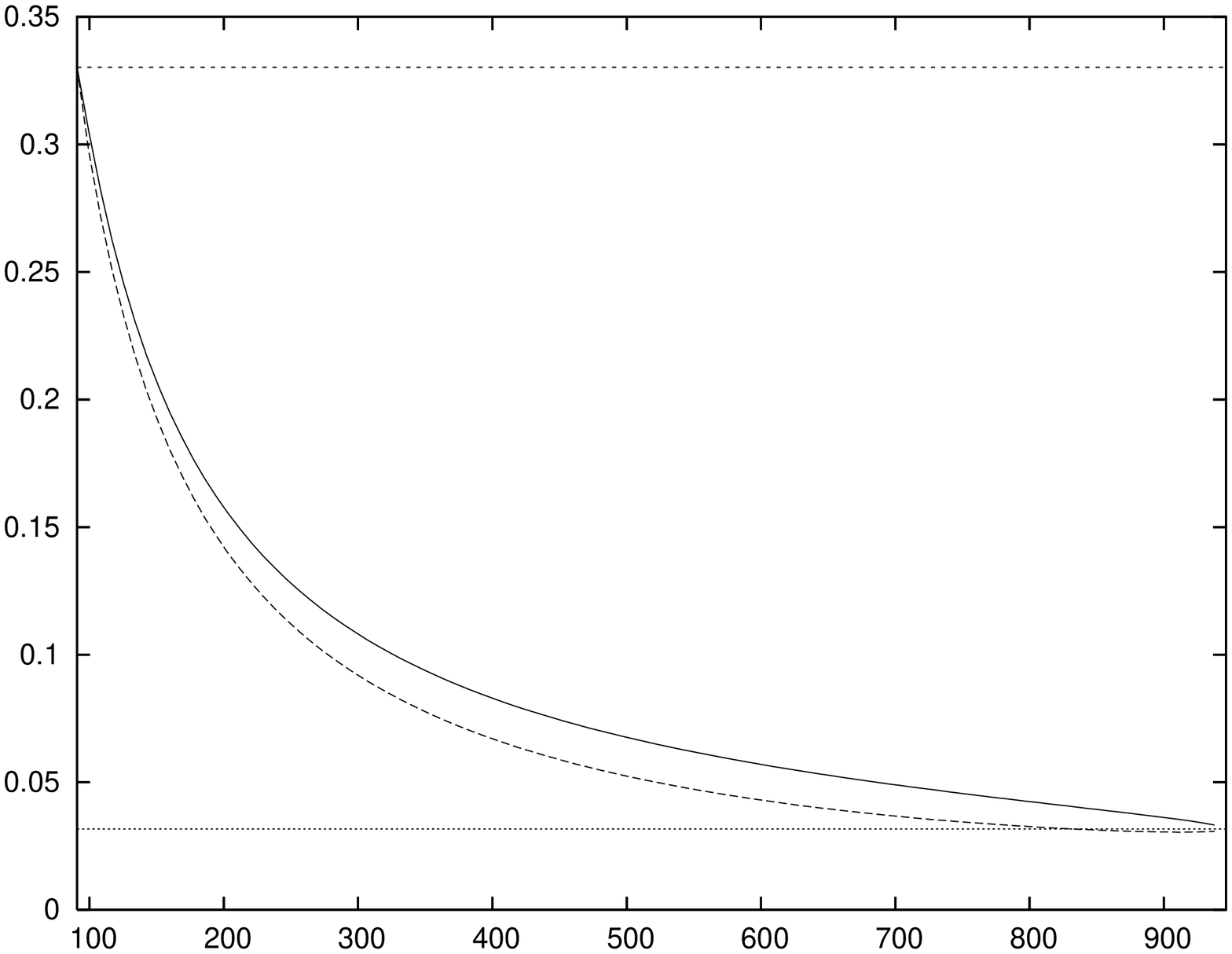}} 
\end{picture}
\vspace{0.2cm}
\caption{On the left, we display the level-crossing in the case of a non-extremal RN-dS black hole,
with $L=1000,M=50,Q=30,\mu=1,e=1,k=1$. The particle and the
black hole have charges with the same sign. One finds $\rp\sim 90.842$,
$\rc \sim 946.214$, $\rmen \sim 9.999$ and $\rzer \sim -1047.056$.
The upper straight line represents $e Q/r_+$, the lower is $e Q/r_c$.
The upper potential is
$\ezp$, the lower one is $\ezm$. Level-crossing occurs, 
but the potential barrier is as large as the
whole spacetime region at hand. On the right, the only change with respect
to the previous figure stays in the smaller value $\mu=0.01$ of the fermion mass.
The upper straight line represents $e Q/r_+$, the lower is $e Q/r_c$.
The upper potential is
$\ezp$, the lower one is $\ezm$. Level-crossing occurs in this case with
a much smaller extent of the potential barrier with respect to $\rc-\rp$.}
\label{fig1}
\end{figure}
\begin{figure}[htbp!]
\begin{picture}(150,150)
\put(-140,160){\includegraphics[height=.30\textheight,
                      width=.24\textheight,angle=-90]{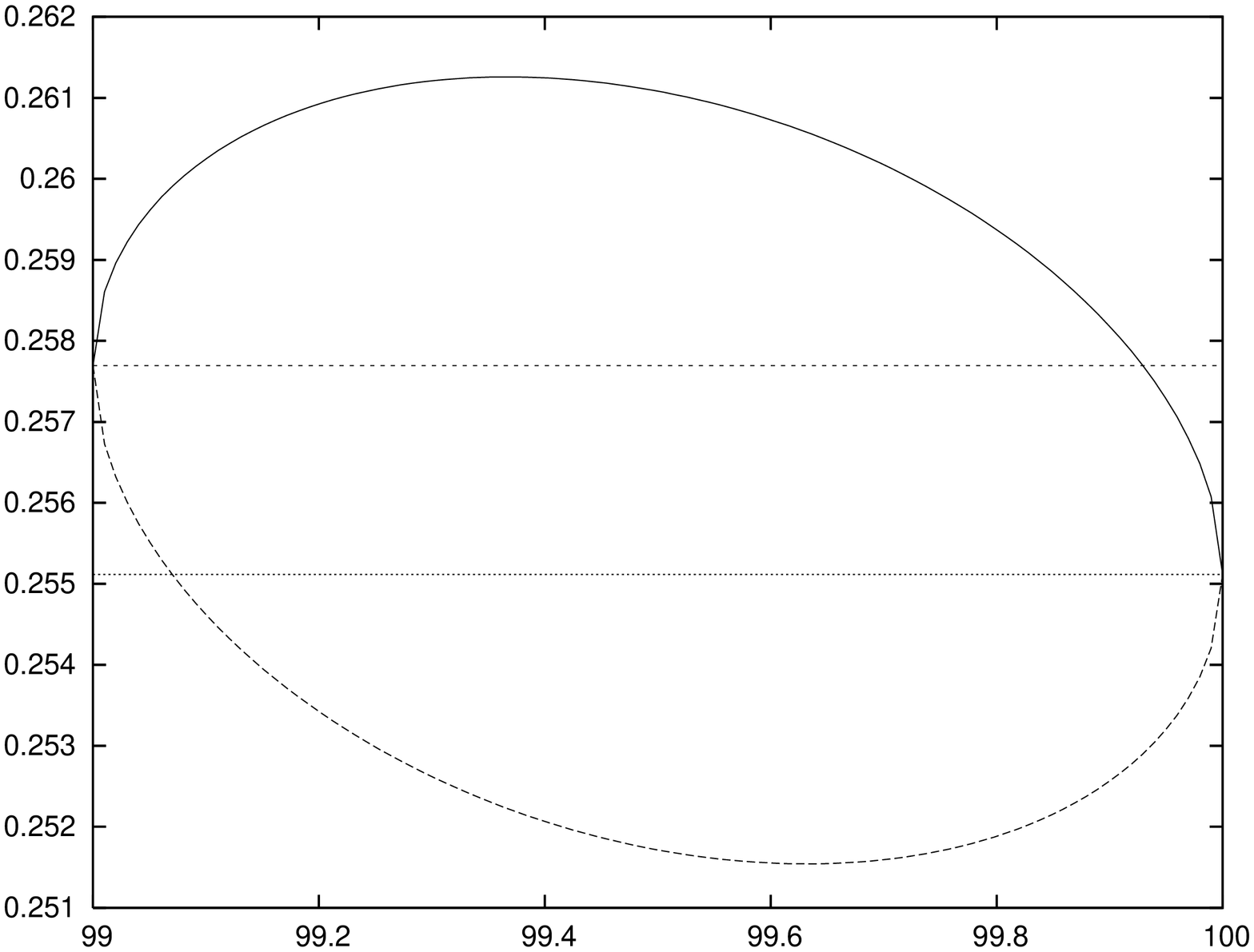}}
\put(90,160){\includegraphics[height=.30\textheight,
                      width=.24\textheight,angle=-90]{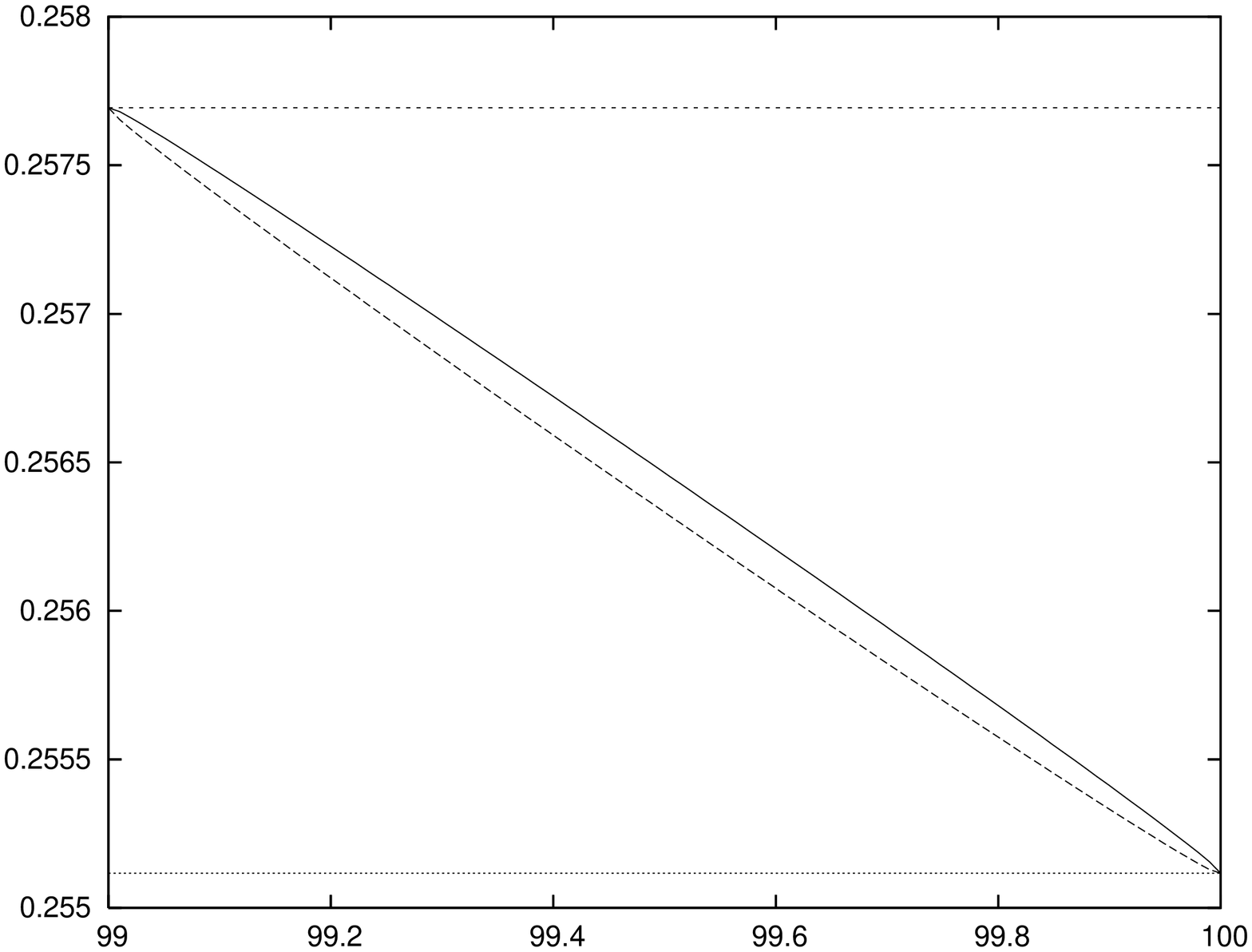}}
\end{picture}
\vspace{0.2cm}
\caption{Level-crossing in the case of a non-extremal RN-dS black hole
having $L=\sqrt{31791}$, $M=1193005/31791$ and $Q=330 \sqrt{190}/\sqrt{31791}$,
which are such that $\rc=100$, $\rp=99$ and $\rmen=10$. Moreover, we choose
$e=1$, $k=1$, and the same sign for the black hole charge and for the particle charge.
The upper straight line represents $e Q/r_+$, the lower is $e Q/r_c$.
The figure on the right displays the potentials for $\mu =1$, and shows that
a very large potential barrier occurs. In the figure on the right one
has $\mu=0.01$ and a very narrower potential barrier.
Note that in the latter case $e Q/r_+ > \mu$ holds, whereas in the former the
opposite inequality is implemented.}
\label{fig2}
\end{figure}
\begin{figure}[htbp!]
\begin{picture}(150,150)
\put(-140,160){\includegraphics[height=.30\textheight,
                      width=.24\textheight,angle=-90]{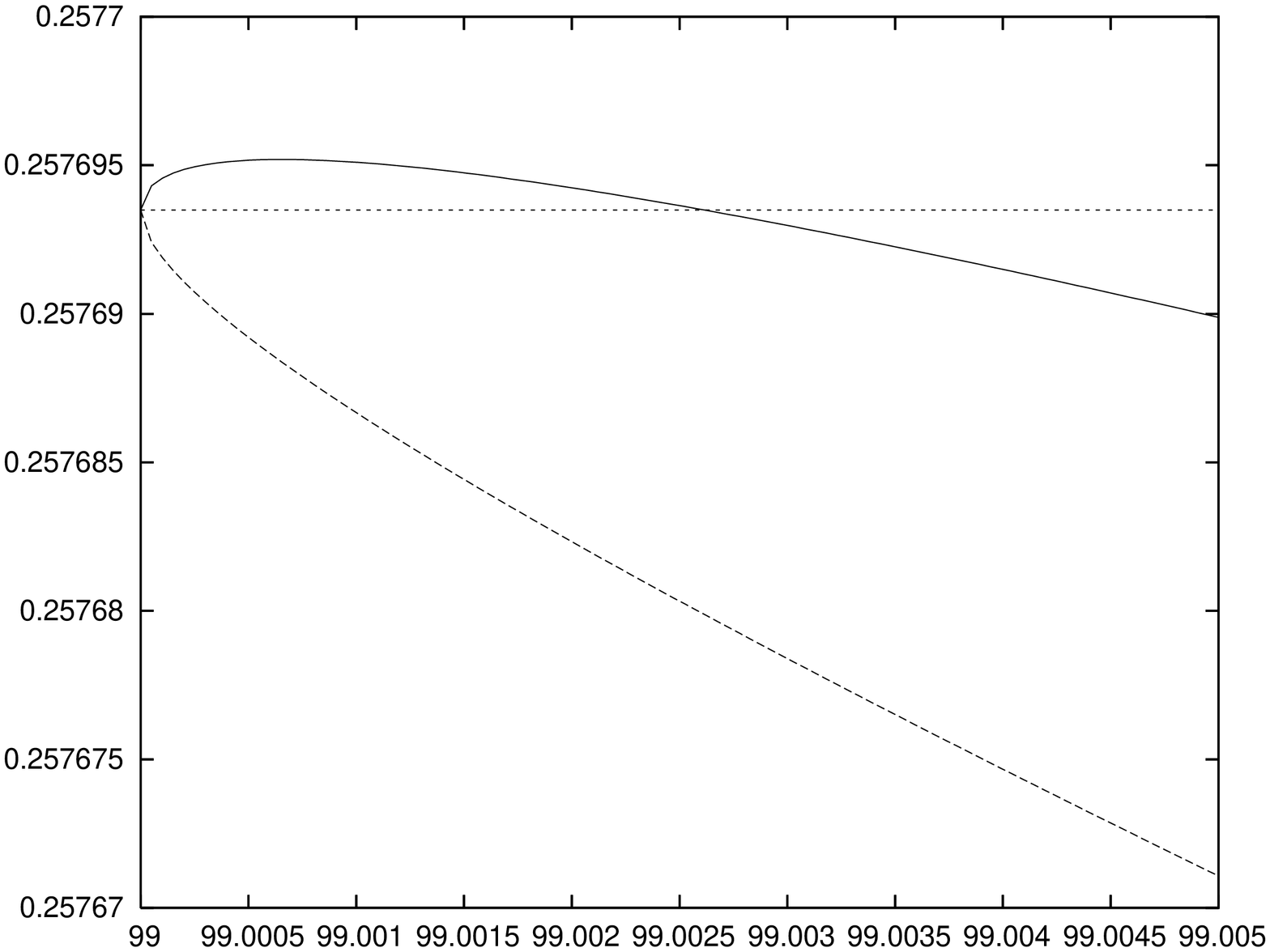}} 
\put(90,160){\includegraphics[height=.30\textheight,
                      width=.24\textheight,angle=-90]{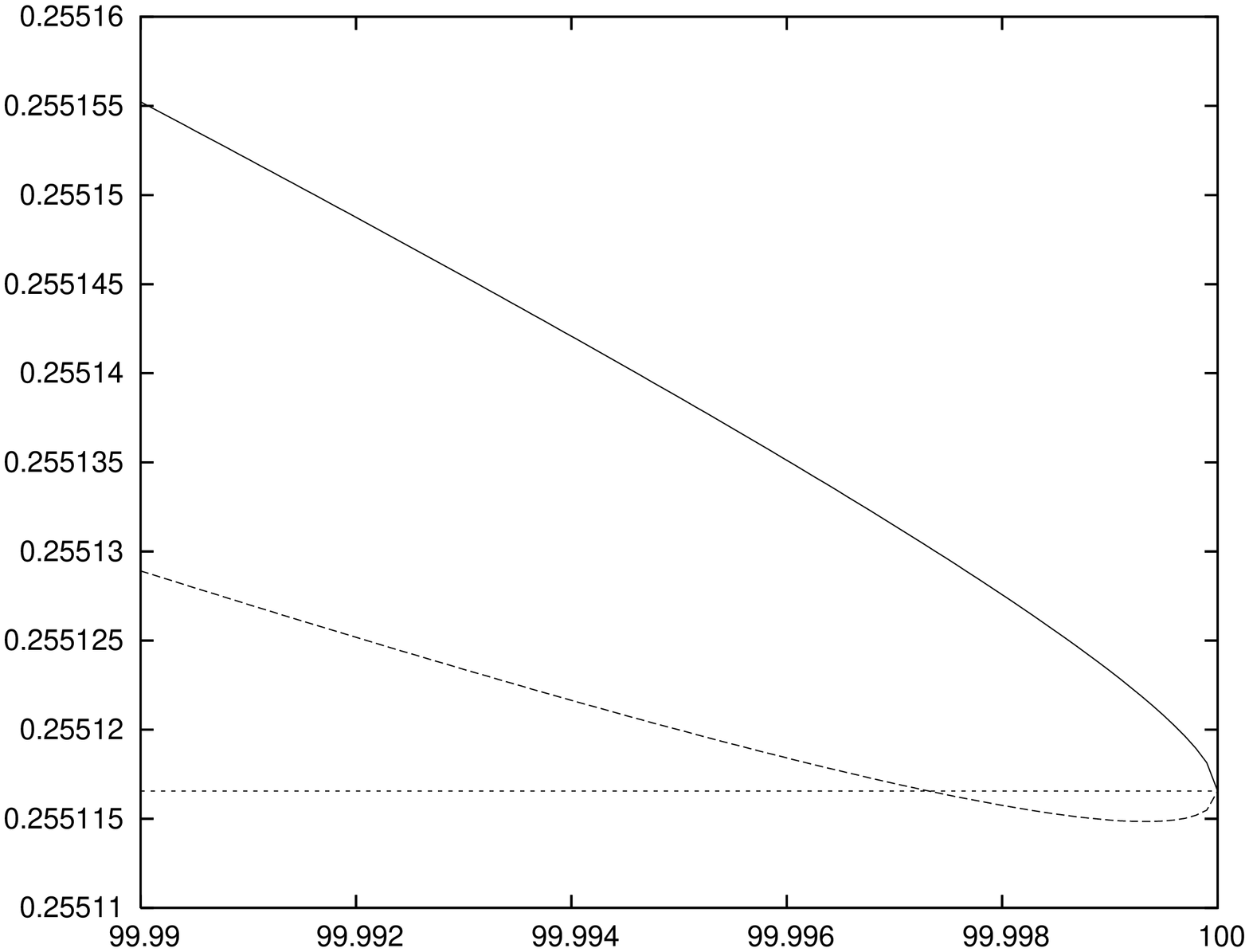}} 
\end{picture}
\vspace{0.2cm}
\caption{Level-crossing in the case of a non-extremal RN-dS black hole
having $L=\sqrt{31791}$, $M=1193005/31791$ and $Q=330 \sqrt{190}/\sqrt{31791}$, as
in the previous figure, 
with $e=1$, $k=1$. We display on the left the presence of a bump near the black hole
horizon in the case of $E_0^+$. Analogously, on the right we show the behavior
of the potentials very near the cosmological horizon.}
\label{fig3}
\end{figure}
\begin{figure}[htbp!]
\begin{picture}(150,150)
\put(-140,160){\includegraphics[height=.30\textheight,
                      width=.24\textheight,angle=-90]{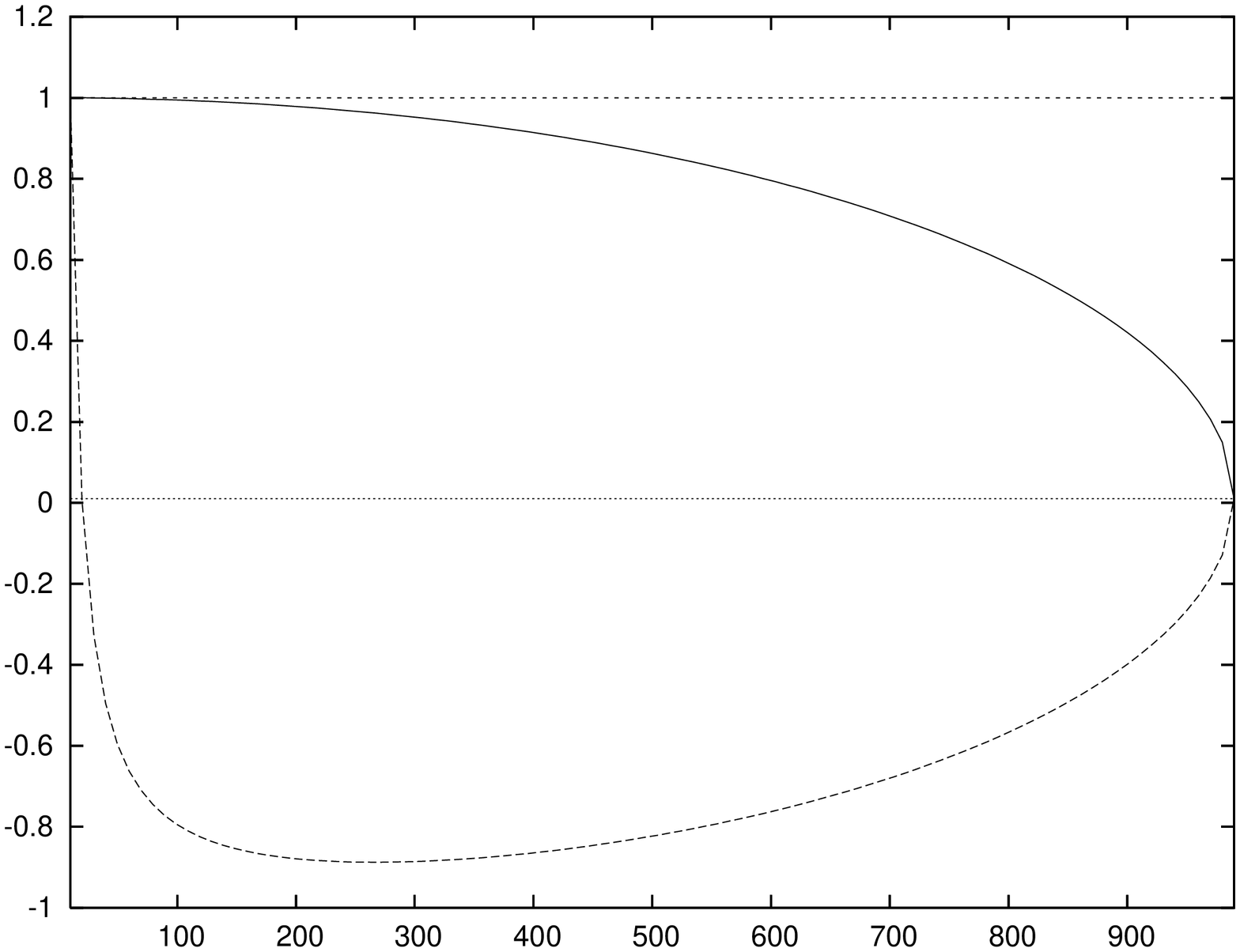}} 
\put(90,160){\includegraphics[height=.30\textheight,
                      width=.24\textheight,angle=-90]{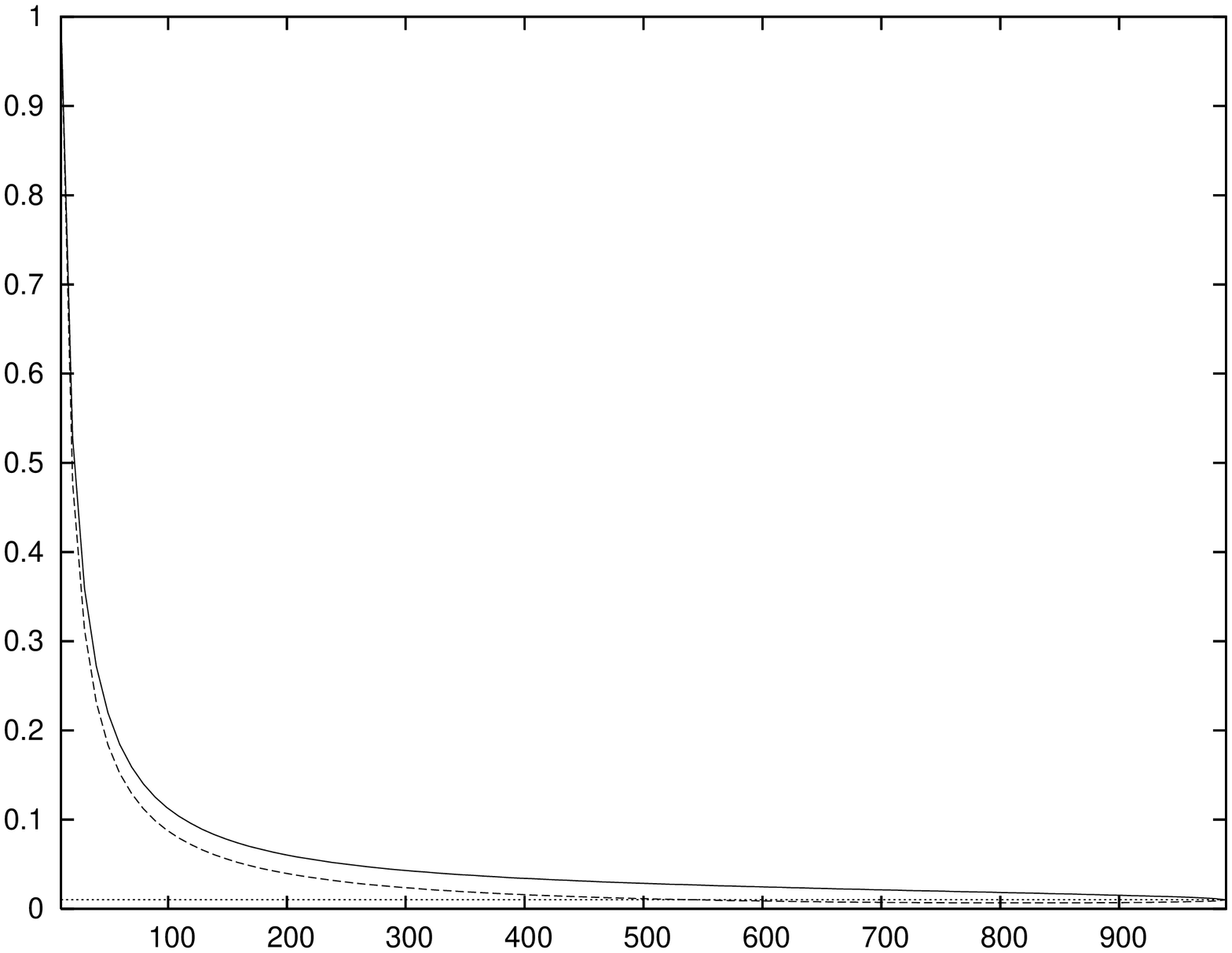}} 
\end{picture}
\vspace{0.2cm}
\caption{Level-crossing in the case of an extremal RN-dS black hole,
with $L=1000,Q=10$ and then $M\simeq 9.999499$ (see eq. (\ref{mextr})).
Particle parameters $e=1,k=1$ are kept fixed, whereas it holds 
$\mu=1$ on the left and $\mu=0.01$ on the right. Level-crossing is more effective
in the latter case, and it occurs
without any bump near the black hole event horizon.}
\label{fig4}
\end{figure}
\begin{figure}[htbp!]
\begin{picture}(150,150)
\put(-140,160){\includegraphics[height=.30\textheight,
                      width=.24\textheight,angle=-90]{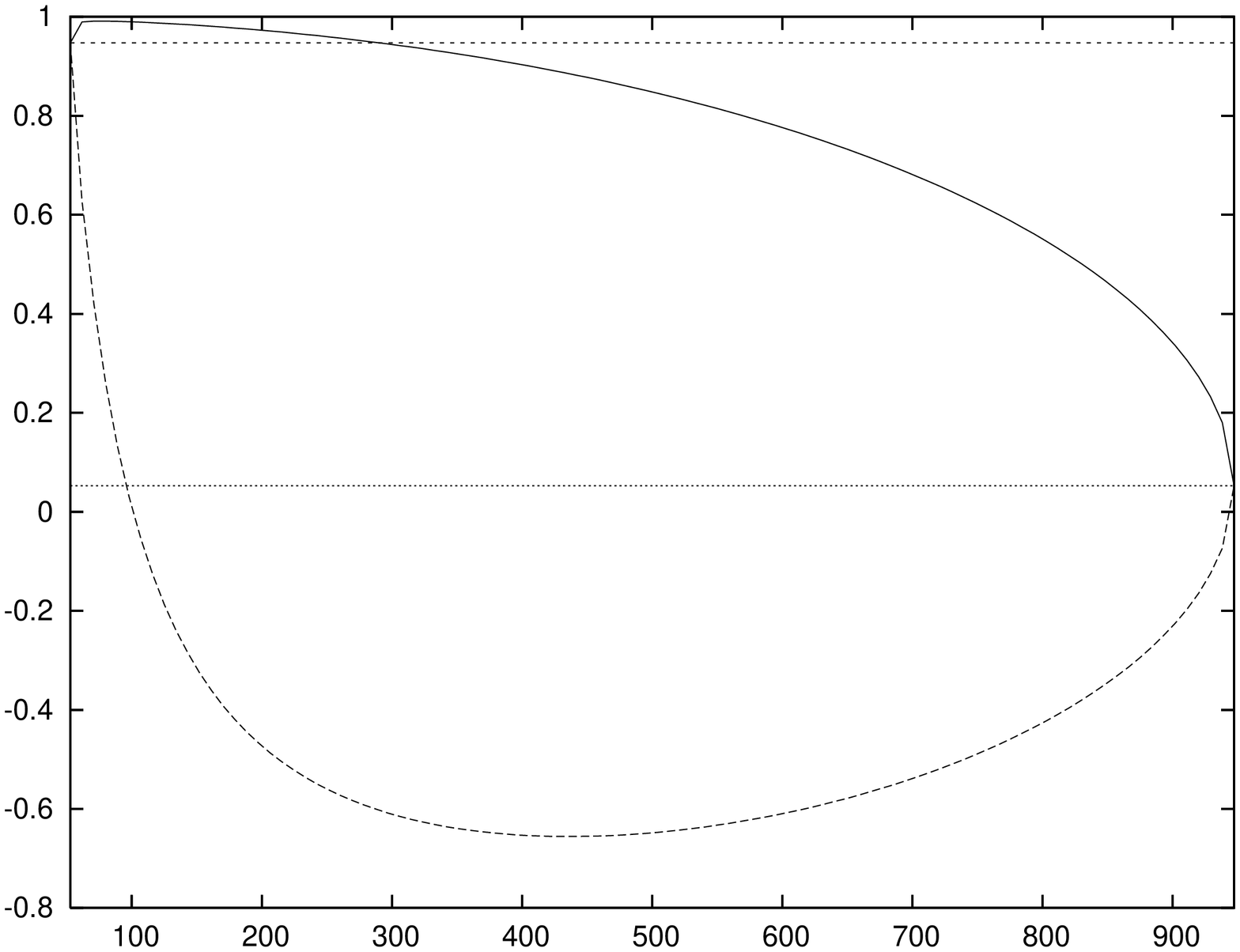}} 
\put(90,160){\includegraphics[height=.30\textheight,
                      width=.24\textheight,angle=-90]{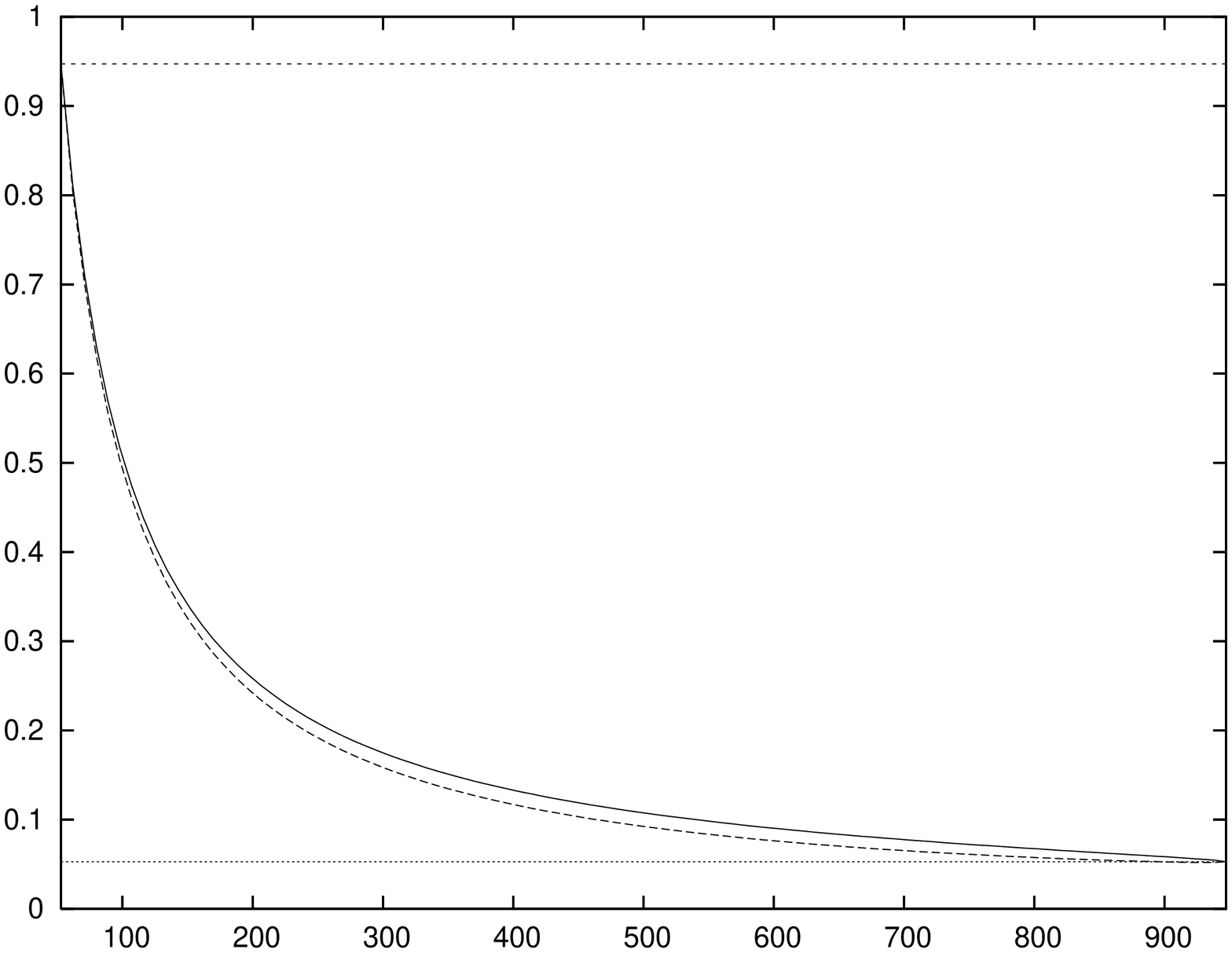}} 
\end{picture}
\vspace{0.2cm}
\caption{Level-crossing in the case of a lukewarm RN-dS black hole,
with $L=1000,M=50,Q=50,e=1,k=1$, and with $\mu=1$ on the left and $\mu=0.01$
on the right. One finds $\rp \sim 52.786$
and $\rc \sim 947.214$.
Level-crossing is qualitatively
similar to the one displayed in figure 1.}
\label{fig5}
\end{figure}
\begin{figure}[h]
\setlength{\unitlength}{1.0mm}
\centerline{\psfig{figure=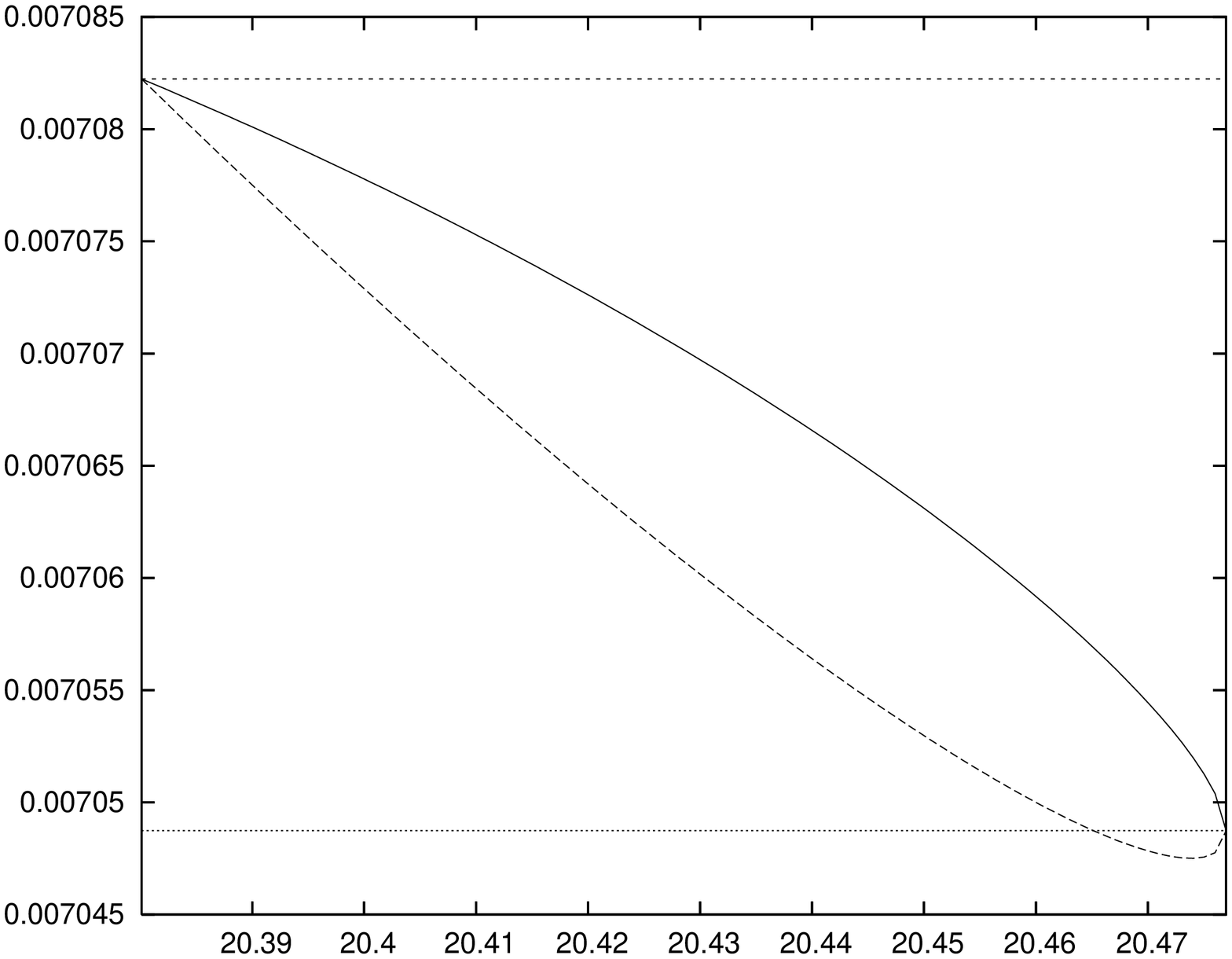,height=8cm,width=5cm,angle=-90}} 
\vspace{0.2cm}
\caption{Level-crossing in the case of an extremal RN-dS black hole,
with $L=50, Q=14.4336845607765725$, which are such that $M=13.608225263871805121$,
$\rp \simeq 20.380115$ and $\rc\simeq 20.476977$. With  $\mu=0.01, e=0.01, k=1$
one gets $\Phi_+ \simeq 0.708 \mu$, which violates (\ref{standard}).}
\label{fig6}
\end{figure}

\noindent
Explicit evaluations of the transmission coefficient which is related to the 
pair-creation phenomenon (discharge) can be given e.g. in a WKB
approximation, as pointed out in the original literature \cite{gibbons,khriplovich,damo}.
See also \cite{belgcaccia} for a short summary. We do not delve into quantitative
evaluation herein in the RN-dS case but limit ourselves to some
estimates in the cases which will be analyzed in the following subsections.

\subsection{the Nariai case}

The potentials $E_0^{\pm} (\chi)$ in the Nariai case are
\beq
E_0^{\pm} (\chi) = e Q \frac{B}{A} \cos (\chi) \pm \sqrt{\frac{\mu^2}{A}+k^2 \frac{B}{A}} \sin (\chi).
\eeq
Also in this case level-crossing is always present, being $E_0^{+} (\pi) < E_0^{-} (0)$ for $e Q>0$
and $E_0^{+} (0) < E_0^{-} (\pi)$ for $e Q<0$. Level-crossing occurs for energies $\omega$ such that  
$E_0^{+} (\pi) \leq \omega \leq  E_0^{-} (0)$ in the former case and for $E_0^{+} (0) \leq \omega \leq  E_0^{-} (\pi)$ 
in the latter one. 
Again, the largeness of the potential barrier
depends on the choice of the parameters. See figure 7.\\

\begin{figure}[htbp!]
\begin{picture}(150,150)
\put(-140,160){\includegraphics[height=.30\textheight,
                      width=.24\textheight,angle=-90]{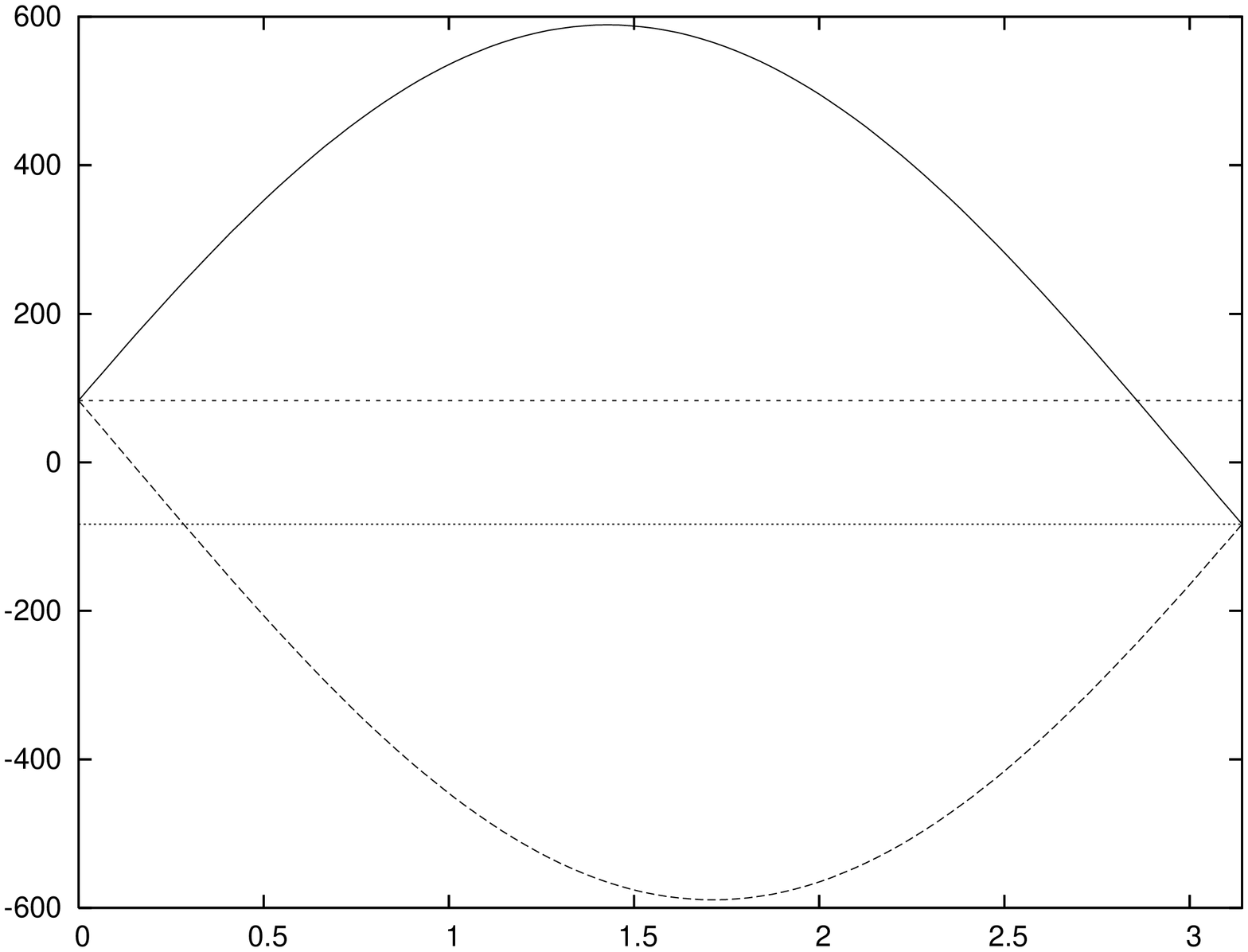}} 
\put(90,160){\includegraphics[height=.30\textheight,
                      width=.24\textheight,angle=-90]{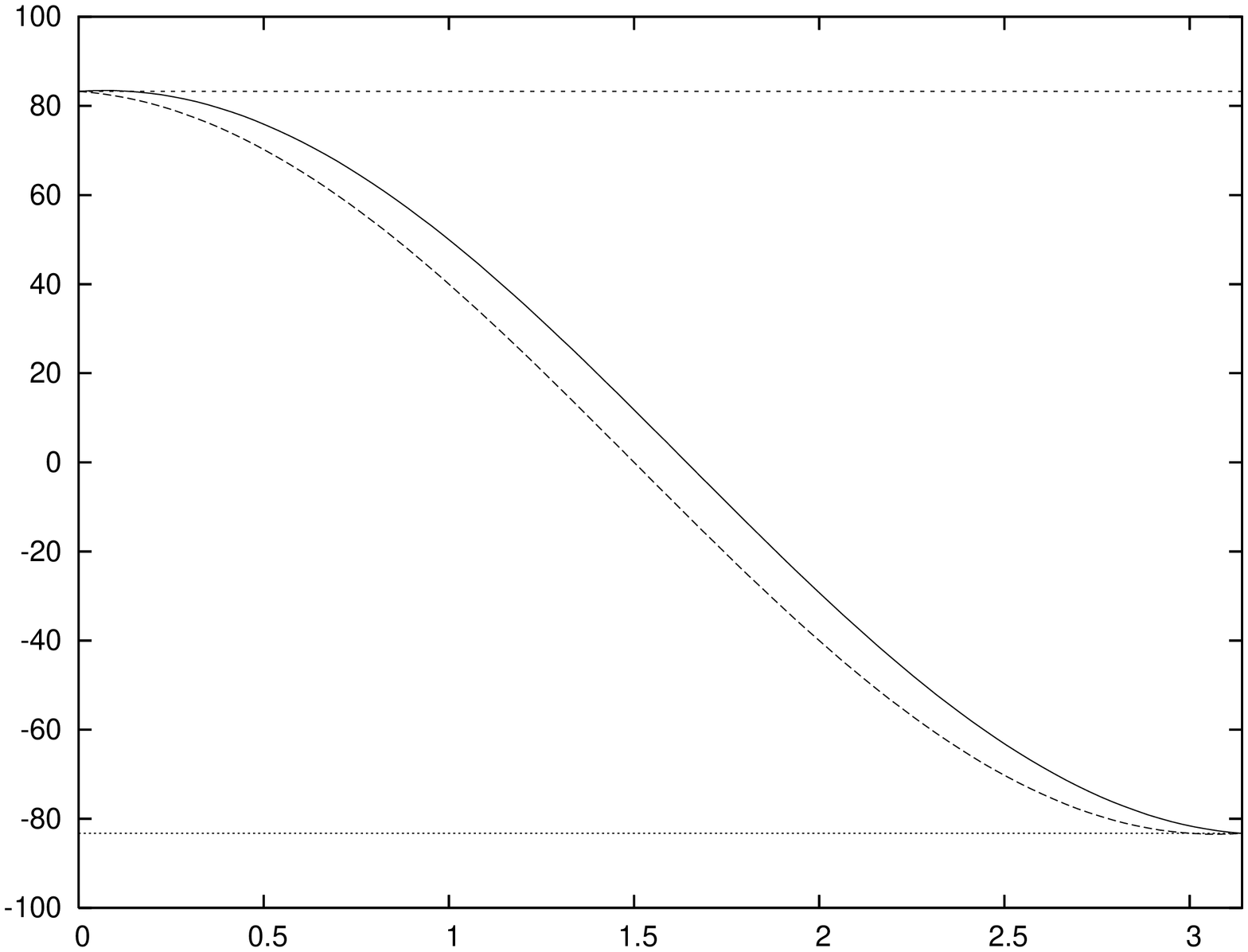}} 
\end{picture}
\vspace{0.2cm}
\caption{Level-crossing in the case of a Nariai solution.
The potentials $E_0^{\pm} (\chi)$ are plotted, with $L=1000, Q=80$. Particle parameters
$e=1,k=1$ are kept fixed; particle mass is chosen to be $\mu=1$ on the left and
$\mu=0.01 $ on the right. Also in the latter case, a bump is present for $E_0^+$ 
near $\chi=0$ and a hollow occurs for $E_0^-$ near $\chi=\pi$.}
\label{fig7}
\end{figure}

An estimation of the transmission coefficient can be given in the WKB approximation;
one obtains \cite{damo}
\beq
|T^{WKB}_{\omega}|^2 = \exp (-2 \int_{\mathrm{barrier}} dx \sqrt{Z_{\omega}}),
\eeq
where $x$ stays for the coordinate defined in (\ref{x-nariai})) and where we have
stressed the dependence on the energy $\omega$ of $T^{WKB}_{\omega}$ and of
\beq
Z_{\omega}=
\left(\frac{B}{A} k^2 +\frac{\mu^2}{A}\right) \sin^2 (\chi (x))-(\omega-e Q \frac{B}{A} \cos (\chi (x)))^2.
\label{z-omega}
\eeq
Let us introduce:
\beq
\mu_k^2 = \frac{B}{A} k^2 +\frac{\mu^2}{A},
\label{mu-k}
\eeq
and also
\beq
E_m = |Q| \frac{B}{A},
\eeq
which corresponds to $\frac{1}{A}$ times the maximum value for the modulus of the
electrostatic field. Note that positivity of $Z_\omega$ requires $\omega^2<\mu_k^2+e^2E_m^2$.
We obtain
\beq
|T_\omega^{WKB}|^2 = \exp \left(-2 \pi |e| E_m (\sqrt{1+\frac{\mu_k^2}{e^2 E_m^2}}-1) \right), \label{indipendente}
\eeq
which does not depend on $\omega$. See Appendix \ref{integral} for more details. 
At the leading order as $\mu_k^2 \ll e^2 E_m^2$ one is also able to recover the
approximation
\beq
|T_\omega^{WKB}|^2 \sim \exp \left(-\pi \frac{\mu_k^2}{|e| E_m}\right),
\eeq
which shares a nice resemblance with the WKB estimate of the 
transmission coefficient related to the pair creation process in a uniform constant electrostatic field. 
Cf. e.g. \cite{damo}.

\subsection{the ultracold cases}

We obtain
\beq
E_I^{\pm} (\chi) = -\frac{e \sqrt{\Lambda}}{2} \chi^2 \pm
\sqrt{2 \Lambda k^2+\mu^2} \chi
\label{uc1-pot}
\eeq
in the case of the first kind of ultracold solution (\ref{ultracold-I}), 
and level-crossing requires that $\omega\leq 0$ for $e>0$ and 
$\omega\geq 0$ for $e<0$.  In the case (\ref{ultracold-II}) one finds 
\beq
E_{II}^{\pm} (x) = -e \sqrt{\Lambda} x \pm
\sqrt{2 \Lambda k^2+\mu^2},
\label{uc2-pot}
\eeq
It is evident that for any $\omega\in \RR$ one obtains level-crossing.   
In figures 8 and 9 level-crossing is displayed for both the ultracold I and the ultracold II cases.

\begin{figure}[htbp!]
\begin{picture}(150,150)
\put(-140,160){\includegraphics[height=.30\textheight,
                      width=.24\textheight,angle=-90]{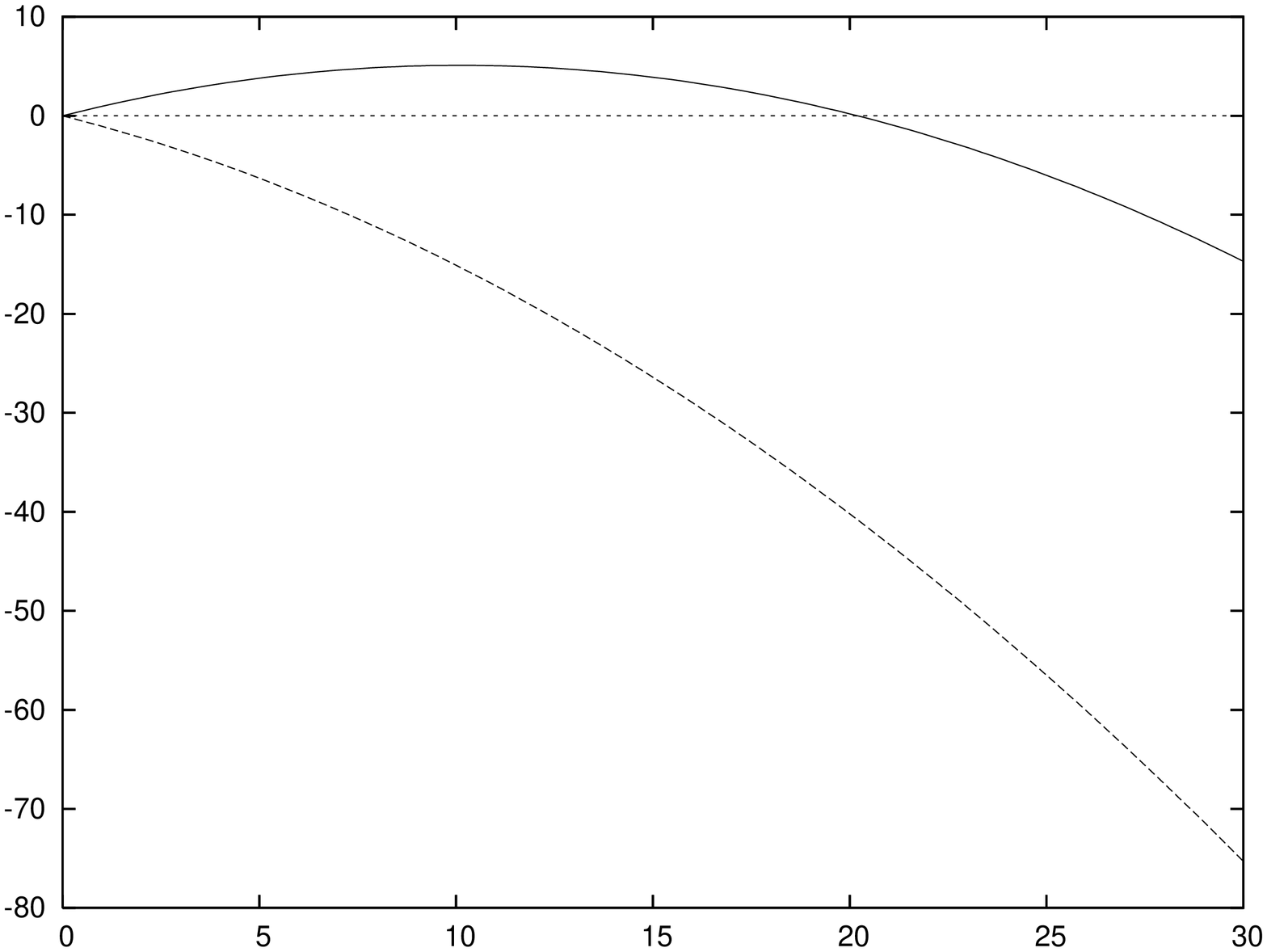}} 
\put(90,160){\includegraphics[height=.30\textheight,
                      width=.24\textheight,angle=-90]{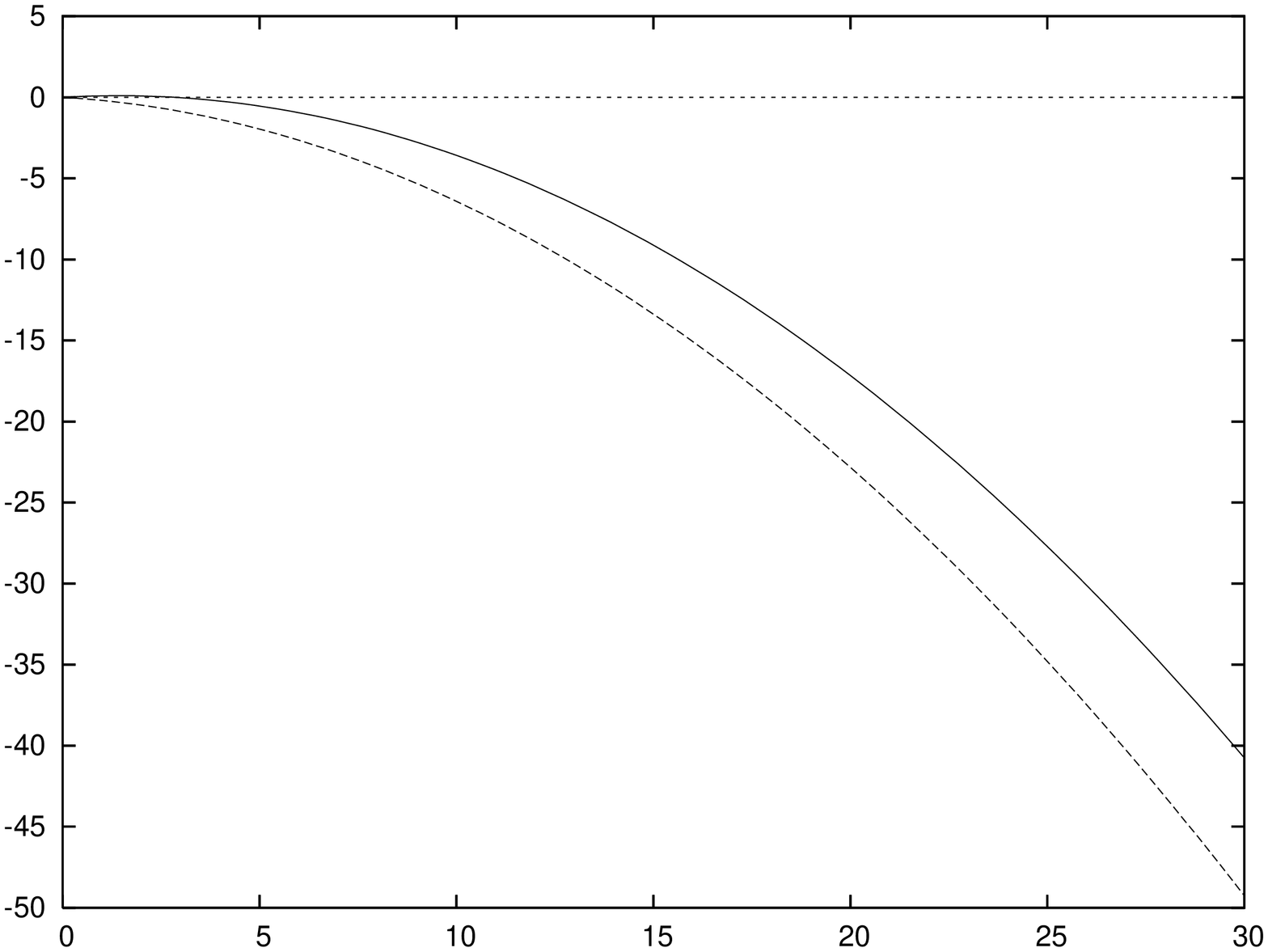}} 
\end{picture}
\vspace{0.2cm}
\caption{Level-crossing in the case of the ultracold I metric. We
choose $\Lambda=0.01,  k=1, e=1$ in both cases and $\mu=1$ on the left 
and $\mu= 0.01$ on the right. We display only
a part of the full plot (but qualitatively all relevant information is given).
The potentials converge both to
zero as $\chi\to 0$ and to $-\infty$ as $\chi\to +\infty$.} 
\label{fig8}
\end{figure}
\begin{figure}[htbp!]
\begin{picture}(150,150)
\put(-140,160){\includegraphics[height=.30\textheight,
                      width=.24\textheight,angle=-90]{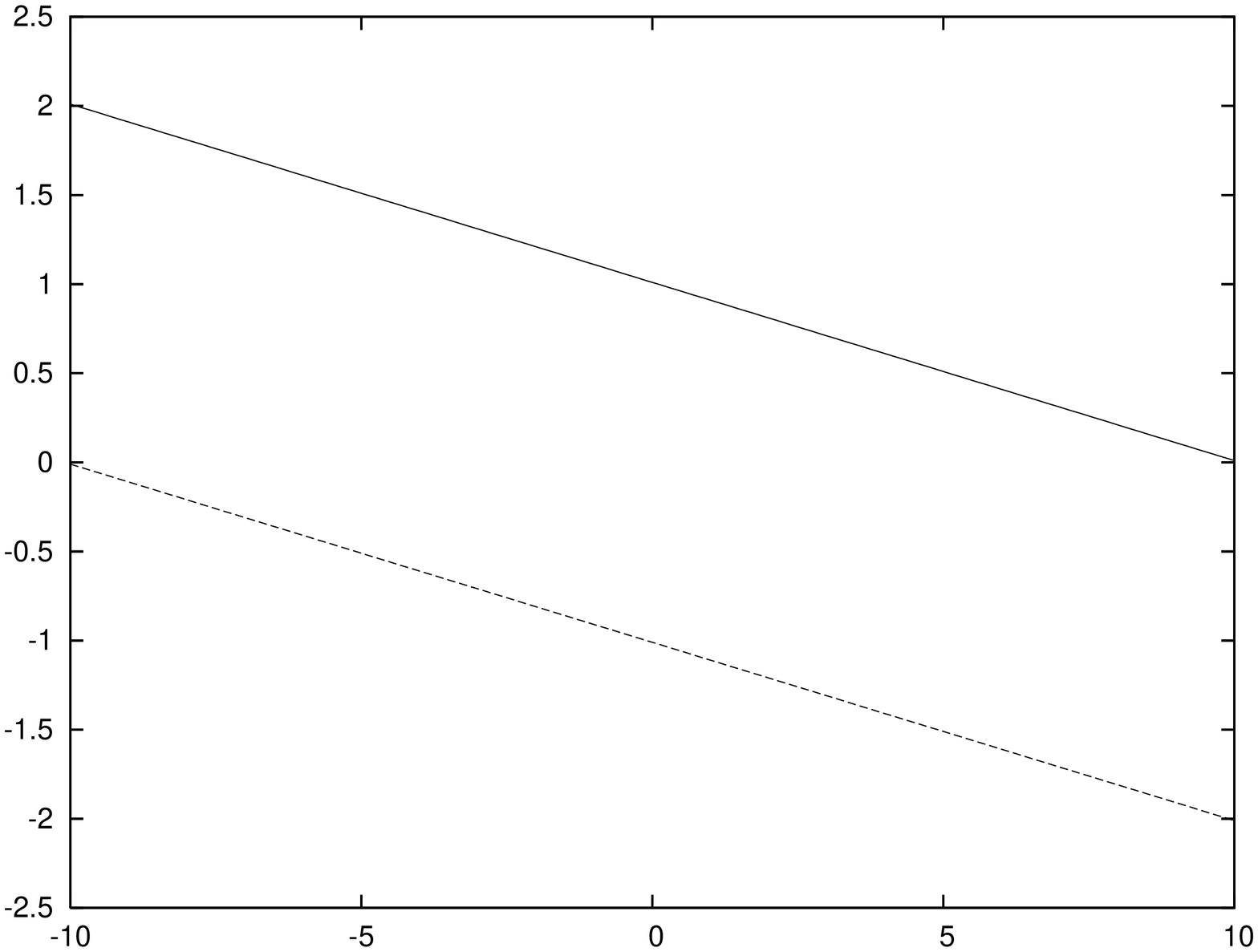}} 
\put(90,160){\includegraphics[height=.30\textheight,
                      width=.24\textheight,angle=-90]{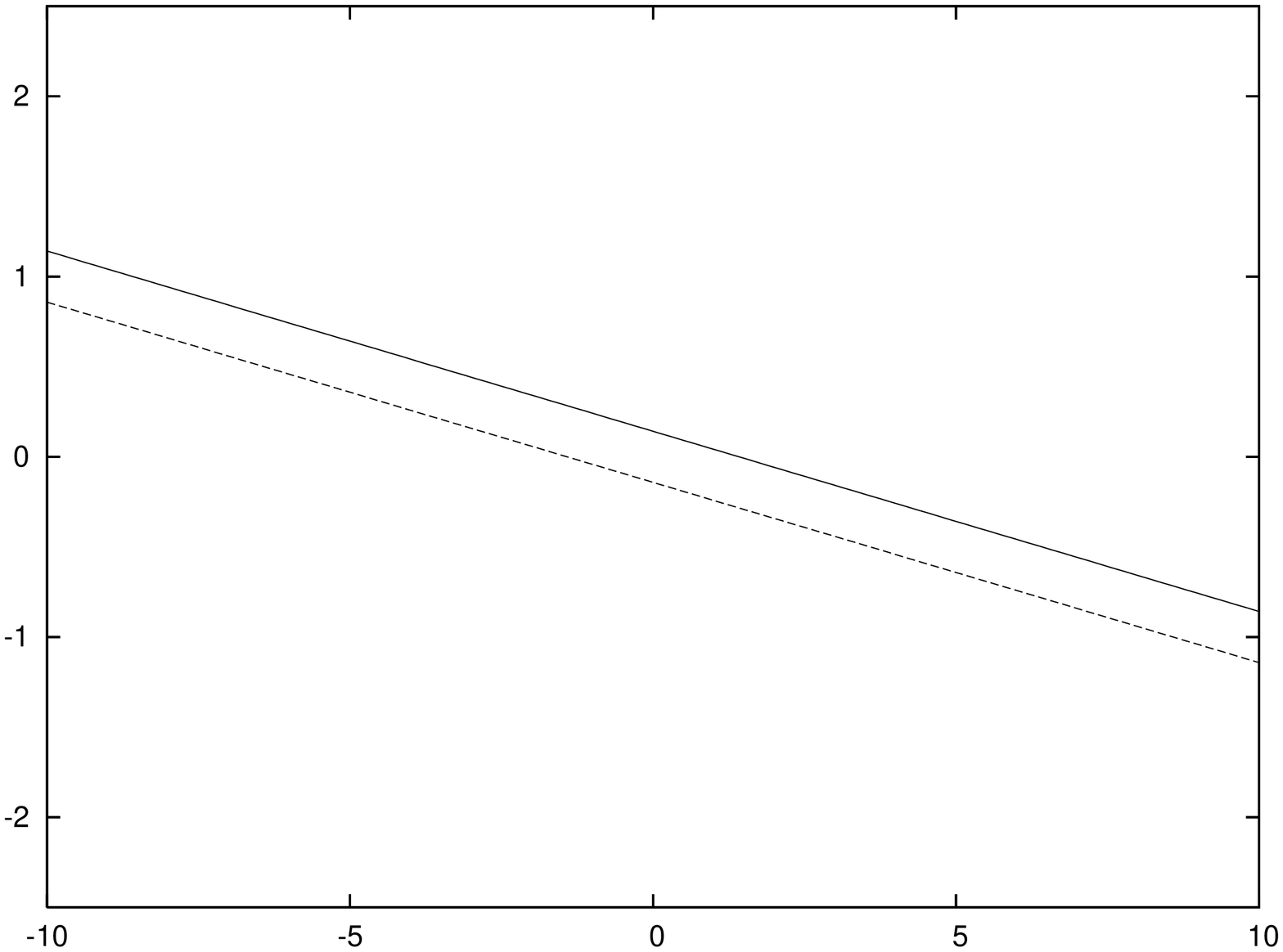}} 
\end{picture}
\vspace{0.2cm}
\caption{Level-crossing in the case of the ultracold II metric. We
choose $\Lambda=0.01, k=1, e=1$ in both cases and $\mu=1$ on the left 
and $\mu= 0.01$ on the right. We display only a part of the full plot,  
which shows of course a linear behavior.}
\label{fig9}
\end{figure}
As to a WKB approximation for $|T_{\omega}|^2$, we get in both cases
\beq
|T_\omega^{WKB}|^2 = \exp \left(-\frac{\pi (\mu^2+2 \Lambda k^2)}{|e| \sqrt{\Lambda}} \right),
\label{traI}
\eeq
which again is independent from $\omega$.
Notice that, by keeping into account that for the electrostatic field
one finds $E=\sqrt{\Lambda}$, and with the replacement $\mu^2\mapsto
\mu^2_k = \mu^2+2 \Lambda k^2$, one obtains again a formula which is very similar to
the one which is associated with the description of the pair creation in a uniform constant electric field in
flat space-time in the same approximation, and this time no requirement about the
smallness of the ratio between $\mu^2_k$ and $|e| E$ is imposed.\\
A deeper analysis for the special cases ultracold II, ultracold I and Nariai is in progress 
\cite{belcadalla}.\\

Some considerations about the problem of the choice of the
quantum state playing the role of vacuum are addressed. 
If one were to assume that 
the positive and negative frequencies associated with the Hamiltonian define the
vacuum, one would end up with the so-called Boulware vacuum, which is viable as 
the real vacuum only in the ultracold II case, where the background temperature is zero \cite{romans}. 
For a Reissner-Nordstr\"om-de-Sitter black hole background, 
a further difficulty arises due to the 
presence of both a cosmological temperature and of a black hole temperature in the
non-extremal and non-lukewarm cases, involving a true non-equilibrium situation. A simpler case is the
extremal one, because of the occurrence of a single temperature for the given manifold, and 
the same considerations can be made in the case of the 
lukewarm solution and in the Nariai case.\\
We are not aware of a rigorous construction for quantum field theory on the given
backgrounds. 
One could expect that, in presence of a single non-zero temperature, 
suitable analyticity requirements
for the fields on the extended manifold can lead to the thermal state as in \cite{gibbons},
and that ``heating up'' the Boulware vacuum (as it can be rigorously done in the
case of a scalar field on a Schwarzschild black hole background \cite{kay}),
taking into account the complication of the
level-crossing displayed above, could be a viable solution. The instability 
associated with the pair-creation process induced by the presence of the electrostatic 
field generated by the black hole still remains, and gives rise to the 
process of discharge we have taken into account. Thermality of the physical state 
modifies such a pair-creation process but the transmission coefficient $|T|$ we have 
calculated for a vacuum situation still plays a role, as it is shown e.g. in Ref. 
\cite{kimleeyoon} for the case of quantum electrodynamics in flat spacetime (see also 
\cite{kim-scalar}). 
One obtains that for an initial thermal state pair-creation is still proportional to 
$|T|^2$ with a multiplicative factor depending on the temperature. We shall 
come back to this topic in \cite{belcadalla}.  
The general RN-dS case is evidently more tricky and challenging, and 
requires a non-equilibrium framework.

\section{Conclusions}

We have shown that, on the background of a charged black hole in de-Sitter space,
massive and charged Dirac particles
are described by an Hamiltonian operator which is well-behaved both on the
cosmological horizon and on the black hole horizon. We have also inferred that
in all cases the point spectrum of the Hamiltonian is empty, and then there is no bound state
and no normalizable time-periodic solution of the Dirac equation. The presence of two
different horizons allows a simpler analysis even in the extremal case.
Moreover, the same occurrence of two
event horizons involving different values of the electrostatic potential
is at the root of the presence in any case of level-crossing
between positive energy states and negative energy ones. This fact {\it per se} is
not enough for claiming that a sensible pair creation effect is present on the given
manifold, due to the fact that a priori the potential barrier to be overcome 
can be very large (even large as almost the whole external manifold in the RN-dS case and in the Nariai one) 
and then the effect is expected to be very suppressed. Nevertheless, in all cases 
examples can be found where the barrier is of much more reduced extent, in such a way 
to allow a physical ground to the pair-creation phenomenon. Some estimates in WKB 
approximation have been given for the transmission coefficient which is related to 
the pair-creation process \cite{damo,khriplovich,gibbons} 
in the case of the Nariai geometry and in the ultracold ones.

\section*{Acknowledgments}
This research was supported in part by Perimeter Institute for Theoretical Physics. 
We thank Francesco Dalla Piazza for his help in drawing figures 10 and 11.

\

\appendix

\section{Absolutely continuous spectrum in the Nariai case}
\label{spe-nariai}

We introduce a decomposition point $\bar{d} \in \RR$ and also the following
self-adjoint operators $H_{-\infty}$ and $ H_{\infty}$ on the respective domains
$D(H_{-\infty})=\{ \vec{g}\in L^2 [(-\infty,\bar{d}],dx]^{2},\; \vec{g}
\hbox{ is locally absolutely continuous}; g_1 (\bar{d})=0;\;
H_{-\infty} \vec{g} \in L^2 [(-\infty,\bar{d}],dx]^{2}\}$, and analogously
$D(H_{\infty})=\{ \vec{g}\in L^2 [[\bar{d},\infty),dx]^{2},\; \vec{g}
\hbox{ is locally absolutely continuous}; g_1 (\bar{d})=0;\;
H_{\infty} \vec{g} \in L^2 [[\bar{d},\infty),dx]^{2}\}$. We define
$P_-:=\lim_{x\to -\infty} P(\chi (x))$ and
$P_+:=\lim_{x\to -\infty} P(\chi (x))$, where the $P(\chi (x))$ is the potential
(\ref{nariai-pot}), and write
\beq
P=P_{\mp}+(P-P_{\mp}).
\label{potdec}
\eeq
The first term on the right hand side  of eq. (\ref{potdec}) is
obviously of bounded variation, whereas the latter term is such that
$|P-P_{+}|\in L^1 [[\bar{d},\infty),dx]$ and
$|P-P_{-}|\in L^1 [(-\infty,\bar{d}],dx]$ respectively. Moreover, notice that
\beq
P_{\mp} = \left[
\begin{array}{cc}
\pm e Q  \frac{B}{A} & 0\cr
0 & \pm e Q  \frac{B}{A}
\end{array}
\right].
\eeq
As a consequence, in both
cases the hypotheses of Theorem 16.7 in \cite{weidmann} are implemented, and one
is allowed to conclude that $H_{-\infty}$ has absolutely continuous spectrum   
in $\RR-\{e Q  \frac{B}{A}\}$, and that $H_{\infty}$ has absolutely continuous spectrum 
in $\RR-\{-e Q  \frac{B}{A}\}$. Then the absolutely continuous spectrum of the self-adjoint extension of the
Hamiltonian operator (\ref{hamiltonian-nariai}) is $\RR$.

\section{Absolutely continuous spectrum in the ultracold I case}
\label{spe-uI}

Let us introduce a self-adjoint extension
$H_{-\infty}$
of the formal differential expression (\ref{hk-ucI}) on the interval $(-\infty,0]$
($0$ is the decomposition point).
Notice that
the potential term in (\ref{hk-ucI}) is
\beq
P(x) = \left[
\begin{array}{cc}
-\frac{e \sqrt{\Lambda}}{2} \exp (2 x) -\mu \exp (x)
& \sqrt{2 \Lambda} k \exp (x)\cr
\sqrt{2 \Lambda} k \exp (x)
&-\frac{e \sqrt{\Lambda}}{2} \exp (2 x) +\mu \exp (x)
\end{array}
\right]
\label{pot-ucI}
\eeq
and it is such that $\lim_{x\to -\infty} P(x)={\mathbb O}$. Moreover,
it is easy to show that $|P(x)|\in L^1 [(-\infty,0],dx]$. As a consequence,
Theorem 16.7 of \cite{weidmann} can be applied and the given self-adjoint extension 
has absolutely continuous spectrum in $\RR-\{0\}$.
It is also true that $0$ is not an eigenvalue for $H_{-\infty}$, because no normalizable
solution exists as a consequence of Levinson theorem (whose applicability is
related to the property that each entry in $P(x)$ is integrable near $x=-\infty$;
cf. \cite{eastham}, p.8). Cf. also \cite{yamada} for the Kerr-Newman case.
Thus $\sigma_{ac}(H_{-\infty})=\RR$. As a consequence (cf. e.g.
\cite{belgcaccia-knds}), also
the absolutely continuous spectrum of the self-adjoint extension of $h_k$ on $\RR$
coincides with the whole real line.

\section{Absolutely continuous spectrum in the ultracold II case}
\label{spe-uII}

Let us notice that the equation $h_k \vec{g} = \lambda
\vec{g}$, by putting $\vec{g} (x) =
\left(\begin{array}{c} w_1 (x) u_1 (x)\cr \frac{1}{w_1 (x)} u_2 (x) \end{array} \right)$,
where $w_1 (x)=\exp (-2 \sqrt{2 \Lambda} k x)$, is equivalent to the following equation:
\beq
\frac{d}{dx} \vec{u} (x) =
\left[
\begin{array}{cc}
0 & (-e \sqrt{\Lambda} x -\alpha) w_1 (x) \cr
(e \sqrt{\Lambda} x +\beta) \frac{1}{w_1 (x)} & 0
\end{array}
\right] \vec{u} (x),
\eeq
where $\alpha=\lambda-\mu$ and $\beta=\lambda+\mu$.
Then, by restricting our attention to the interval $[c,\infty)$, with $c>0$, and by applying
Theorem 2 in \cite{hs-ac} the result follows. Let us define, according to the
notations of \cite{hs-ac},
$p_1=(-e \sqrt{\Lambda} x +\mu) \exp (-2 \sqrt{2 \Lambda} k x)=:p_{11}+p_{12}$,
$p_2=(-e \sqrt{\Lambda} x -\mu) \exp ( 2 \sqrt{2 \Lambda} k x)=:p_{21}+p_{22}$,
$\alpha_1 =\exp (-2 \sqrt{2 \Lambda} k x)$ and
$\alpha_2 =\exp ( 2 \sqrt{2 \Lambda} k x)$; $p_{11}:=-e \sqrt{\Lambda} x
\exp (-2 \sqrt{2 \Lambda} k x)$ and $p_{21}:= -e \sqrt{\Lambda} x
\exp ( 2 \sqrt{2 \Lambda} k x)$. One obtains $p_{11} p_{21}= e^2 \Lambda x^2>0$
and
$\int_c^{+\infty} dx \sqrt{p_{11} p_{21}}=+\infty$. Moreover
both $\int_c^{+\infty} dx \alpha_1 \sqrt{\frac{p_{21}}{p_{11}}}$ and
$\int_c^{+\infty} dx \alpha_2 \sqrt{\frac{p_{11}}{p_{21}}}$ diverge (it is
sufficient that one of them diverges \cite{hs-ac}). Moreover,
if $\eta = (\frac{p_{21}}{p_{11}})^{1/4}$, then $\Delta:= \frac{d\eta}{dx}
\frac{1}{\eta \sqrt{p_{11} p_{21}}}=\frac{\sqrt{2 \Lambda} |k|}{|e| x}$ is such that
$\lim_{x\to +\infty} \Delta = 0$ and $\frac{d\Delta}{dx}\in L^1 [[c,+\infty),dx]$. Moreover,
$\frac{\alpha_1}{p_{11}}, \frac{\alpha_2}{p_{21}},
\frac{p_{12}}{p_{11}}, \frac{p_{22}}{p_{21}}$ are long-range (in the sense that
they vanish as $x\to +\infty$ and their derivative is in $L^1 [[c,+\infty),dx]$).
Then the hypotheses of Theorem 2 in \cite{hs-ac} are implemented, which means that
the absolutely continuous spectrum is $\RR$.\\

\section{Evaluation of the Nariai transmission integral}
\label{integral}
\begin{figure}[h]
\setlength{\unitlength}{1.0mm}
\centerline{\psfig{figure=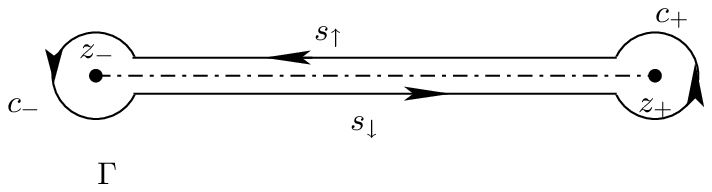,width=5cm,angle=0}} 
\vspace{0.2cm}
\caption{Circuit $\Gamma$ around the cut in the Riemann sheet of $f$} \label{fig:imm1}
\end{figure}

We need to evaluate
\begin{eqnarray}
\int_{\rm barrier} \sqrt {Z_\omega} \ dx 
\end{eqnarray}
where $ Z_\omega$ is given in (\ref{z-omega}) and $\chi(x)=2\arctan e^x$ (cf. (\ref{x-nariai})). 
Using
\begin{eqnarray*}
&& \cos \chi(x)= -\tanh x,\\
&& \sin \chi(x)=\frac 1{\cosh x},
\end{eqnarray*}
we can rewrite
\begin{eqnarray}
&& \int_{\rm barrier} \sqrt {Z_\omega} \ dx=\int_{\rm barrier} \sqrt {\left( \frac BA k^2 +\frac {\mu^2}A \right)-
\left(\omega \cosh x +eQ \frac BA \sinh x \right)^2} \ \frac {dx}{\cosh x} \cr
&& \qquad\qquad\quad\ =\int_{\rm barrier} \sqrt {\left( \frac BA k^2 +\frac {\mu^2}A \right)e^{-2x} -\frac 14 \left( [\omega+eQ \frac BA]+
[\omega -eQ \frac BA]e^{-2x} \right)^2} \ \frac {2e^{2x}}{1+e^{2x}} \ dx.
\end{eqnarray}
If we define $\mu_k^2$ as in (\ref{mu-k}) and $\omega_{\pm} =\omega\pm eQ\frac BA$, and change variable to $z=e^{2x}$ we get
\begin{eqnarray}
&& \int_{\rm barrier} \sqrt {Z_\omega} \ dx=\int_{z_-}^{z_+} dz\ \frac 1{1+z}\ \sqrt {-\frac {\omega_-^2}4 \frac 1{z^2}+\left( \mu_k^2 -\frac 12
\omega_+ \omega_- \right)\frac 1z -\frac {\omega_+^2}4},
\end{eqnarray}
where $0\leq z_- < z_+$ are the two real solutions of
$$
-\frac {\omega_-^2}4+\left( \mu_k^2 -\frac 12 \omega_+ \omega_- \right)z -\frac {\omega_+^2}4 \ z^2 =0.
$$
Note that such solutions exist if the discriminant of the polynomial is positive. This gives
$$
0< \mu_k^2 (\mu_k^2 -\omega_+ \omega_-)=\mu_k^2+e^2 Q^2 \frac {B^2}{A^2}-\omega^2.
$$
The integral is indeed well defined for $-\sqrt {\mu_k^2+e^2 Q^2 \frac {B^2}{A^2}} < \omega <\sqrt {\mu_k^2+e^2 Q^2 \frac {B^2}{A^2}}$.
Note however that,  
only the region of level crossing 
$-|eQB/A|\leq \omega \leq |eQB/A|$, corresponding to $\omega_+ \omega_- \leq0$, is relevant for computing the transmission coefficient. 
Let us first distinguish the ``generic case'' $\omega_+ \omega_-\neq 0$
from the ``particular case'' $\omega_+ \omega_-=0$.\\
In order to compute the integral in the generic case, the residue method is used. Let us cut the complex plane $\mathbb {C}$ along the segment
$[z_-,z_+]\subset \mathbb {R}_{+}$ (note that $z_->0$ in this case), so defining a Riemann sheet for the square root
$f(z)=\sqrt {(z_+-z)(z-z_-)\omega_+^2/4}$. In particular we choose the phase of $(z_+-z)(z-z_-)\omega_+^2/4$ to be $0$ (modulo $4\pi$) along the lower
border of the cut, so that it will be $2\pi$ (modulo $4\pi$) along the upper border. Thus, it makes sense to take the closed path
$\Gamma =s_{\downarrow} \cup c_+ \cup s_{\uparrow} \cup c_-$ as in figure \ref{fig:imm1}.

\begin{figure}[h]
\setlength{\unitlength}{1.0mm}
\centerline{\psfig{figure=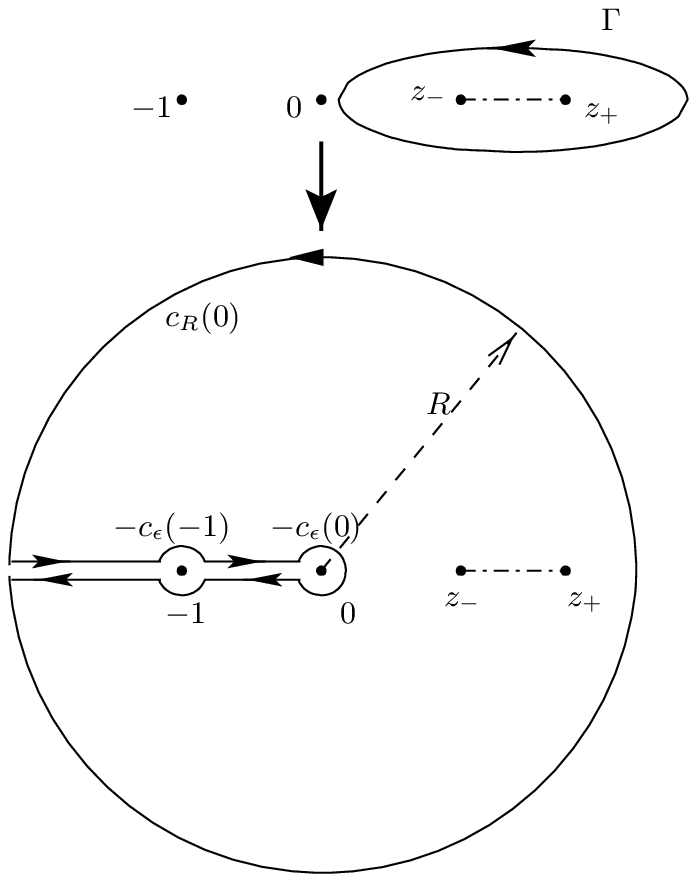,width=5cm,angle=0}} 
\vspace{0.2cm}
\caption{Blow up of the path $\Gamma$} \label{fig:imm2}
\end{figure}

When the radii of the two circles $c_\pm$ are set to zero, $\Gamma$
approaches the cut without crossing any singularity so that the value of the integral
\begin{eqnarray}
I=\oint_{\Gamma} dz \ \frac 1{1+z} \frac {f(z)}z
\end{eqnarray}
does not change. In this limit, the contributions from the circles vanish, whereas
$$
\int_{s_{\downarrow}} \longrightarrow \int_{z_-}^{z^+}, \qquad\quad\ \int_{s_{\uparrow}} \longrightarrow -\int_{z_-}^{z^+}
$$
and being the phase of $f$ equal to $0$ on the lower cut and to $\pi$ on the upper cut, we see that
\begin{eqnarray}
\int_{\rm barrier} \sqrt {Z_\omega} \ dx=\int_{z_-}^{z^+} dz \ \frac 1{1+z} \frac {f(z)}z =\frac 12 I.
\end{eqnarray}
To compute the integral $I$, let us blow up $\Gamma$, without crossing the poles of the integrand (which are $z=0, -1 , \infty$) like in figure
\ref{fig:imm2}.

We see that (cf. figure (\ref{fig:imm2}))
\beq
I=-\oint_{c_\epsilon (0)} dz \ \frac 1{1+z} \frac {f(z)}z -\oint_{c_{\epsilon} (-1)} dz \ \frac 1{1+z} \frac {f(z)}z
+\oint_{c_R (0)} dz \ \frac 1{1+z} \frac {f(z)}z
\eeq
where we used $c_r(z)$ to indicate the counterclockwise oriented circle with center $z$ and radius $r$.
Note that if we take the change of variable $z=1/t$ in the last integral, we have \footnote{Orientation changes because $\arg (1/z)=-\arg z$}
$c_R(0)\longrightarrow -c_{1/R} (0)$ so that
\begin{eqnarray}
&& I=-\oint_{c_\epsilon (0)} dz \ \frac 1{1+z} \frac {f(z)}z -\oint_{c_{\epsilon} (-1)} dz \ \frac 1{1+z} \frac {f(z)}z
+\oint_{c_{1/R} (0)} dt \ \frac 1{1+t} f(1/t)\cr
&& \quad =-2\pi i \left( {\rm res}_{z=0} \left( \frac 1{1+z} \frac {f(z)}z \right)
+{\rm res}_{z=-1} \left( \frac 1{1+z} \frac {f(z)}z \right) -{\rm res}_{z=0} \left( \frac 1{1+z} f(1/z) \right) \right)\cr
&& \quad =-2\pi i \left( f(0)-f(-1) -\lim_{z\rightarrow 0} (zf(1/z)) \right)\cr
&& \quad =2\pi i \left(-\sqrt {-\frac {\omega_-^2}4}+\sqrt{-(\mu_k^2+\frac 14 (\omega_+-\omega_-)^2)}+
\sqrt {-\frac {\omega_+^2}4}  \right).
\end{eqnarray}
To compute the square roots we need to specify their phases. This is easily done looking at the Riemann sheet.
Indeed, if we look at the real axis, on $z<z_-$ the phase of $f^2$ is $3\pi$ (modulo $4\pi$) so that $f(x)=-i|f(x)|$ for $x=0, -1$.
For the last root, we note that
$$
\lim_{z\rightarrow 0} zf(1/z)=\lim_{z\rightarrow \infty} f(z)/z
$$
and because the monodromy of $z=\infty$ is trivial (the phase of $f(z)$ does not change (modulo $2\pi$) if $|z|$ is very large
and ${\rm arg} (z)$ varies by a period), $z=\infty$ is indeed a pole (and not a branch point) and this limit does not depend
on the phase of $z$. Thus, we can compute it along the positive real axis. But there, the phase of $f(z)^2$, for $z>z_+$ is $\pi$, so that
$f(z)=i|f(z)|$, and finally we have
\begin{eqnarray}
&& I=2\pi i \left( -\left(-i \frac {|\omega_-|}2\right)+\left(-i\sqrt{\mu_k^2+\frac 14 (\omega_+-\omega_-)^2}\right) +i\frac {|\omega_+|}2 \right)\cr
&& \quad = 2\pi \left( \sqrt{\mu_k^2+\frac 14 (\omega_+-\omega_-)^2} -\frac 12 (|\omega_+|+|\omega_-|) \right).
\end{eqnarray}
Note that in the physically interesting case, that is when $\omega_+ \omega_-<0$, we have $|\omega_+|+|\omega_-|=|\omega_+-\omega_-|$
which does not depend on $\omega$, and reproduces exactly (\ref{indipendente}).\\
It remains only to check the particular cases, which are however easily obtained by direct integration. 
For example, for $\omega_-=0$, so that $\omega_+=2eQB/A$ we have
\beq
I/2= \int_{\rm barrier} \sqrt {Z_\omega} \ dx =\frac 12 \int_0^{4\frac {\mu_k^2}{\omega_+^2}} \frac {dz}{1+z} \sqrt {\frac 4z \mu_k^2 -\omega_+^2}.
\label{p+}
\eeq
Introducing the new integration variable $s^2= \frac 4z \mu_k^2 -\omega_+^2$ we get
\begin{eqnarray*}
&& I/2= \int_0^\infty \left[\frac 1{\omega_+^2+s^2} -\frac 1{4\mu_k^2 +\omega_+^2 +s^2}\right]\ s^2 ds
=\int_0^\infty \left[-\frac {\omega_+^2}{\omega_+^2+s^2} +\frac {4\mu_k^2 +\omega_+^2}{4\mu_k^2 +\omega_+^2 +s^2}\right]\ ds \cr
&& \qquad =-\frac \pi2 |\omega_+| +\pi\sqrt {\mu_k^2+\frac 14 \omega_+^2}\ .
\end{eqnarray*}
For $\omega_+=0$ ($\omega_-=-2eQB/A$) we have
\begin{eqnarray*}
&& I/2=\int_{\frac {\omega_-^2}{4\mu_k^2}}^\infty \frac 1{z(1+z)} \sqrt {\mu_k^2 z-\frac {\omega_-^2}4 }
=\frac 12 \int_0^{4\frac {\mu_k^2}{\omega_-^2}} \frac {dt}{1+t} \sqrt {\frac 4t \mu_k^2 -\omega_-^2},
\end{eqnarray*}
where we used the change of variable $t=1/z$, obtaining the same integral as in (\ref{p+}) with $\omega_-$ in place of $\omega_+$,
giving thus the same result, being $|\omega_\pm|=2|eQB/A|$.


\end{document}